\documentclass[nonacm,sigconf,screen]{acmart}

\usepackage{amsmath}
\usepackage{float}
\usepackage{tablefootnote}
\usepackage{url}

\floatstyle{plaintop}
\restylefloat{table}

\newcommand{\figureref}[1]{Figure~\ref{#1}}

\newcommand{\labeltitle}[1]{\vskip 0.03in \noindent\textbf{#1}}
\newcommand{\sectionref}[1]{Section~\ref{#1}}
\newcommand{\sectionsref}[2]{Sections~\ref{#1}~and~\ref{#2}}

\newcommand{\tableref}[1]{Table~\ref{#1}}
\newcommand{\tablesref}[2]{Tables~\ref{#1}~and~\ref{#2}}
\newcommand{\eg}{e.\,g.,\ }

\definecolor{red}{RGB}{221, 33, 24}
\definecolor{green}{RGB}{29, 93, 33}
\definecolor{blue}{RGB}{51, 114, 171}
\definecolor{orange}{RGB}{255, 128, 47}
\definecolor{purple}{RGB}{147, 105, 185}

\begin{document}

\title[An Empirical Evaluation of Serverless Cloud Infrastructure for Large-Scale Data Processing]{An Empirical Evaluation of Serverless Cloud Infrastructure for Large-Scale Data Processing}

\author{Thomas Bodner}
\affiliation{
  \institution{HPI, U Potsdam}
  \country{}
  }
\author{Theo Radig}
\affiliation{
  \institution{HPI, U Potsdam}
  \country{}
  }

\author{David Justen}
\affiliation{
  \institution{BIFOLD, TU Berlin}
  \country{}
  } 

\author{Daniel Ritter}
\affiliation{
  \institution{SAP}
  \country{}
  }

\author{Tilmann Rabl}
\affiliation{
  \institution{HPI, U Potsdam}
  \country{}
  }  

\begin{abstract}
Data processing systems are increasingly deployed in the cloud. While monolithic systems run fully on virtual servers, recent systems embrace cloud infrastructure and utilize the disaggregation of compute and storage to scale them independently. The introduction of serverless compute services, such as AWS Lambda, enables finer-grained and elastic scalability within these systems. Prior work shows the viability of serverless infrastructure for scalable data processing, yet also sees limitations due to variable performance and cost overhead, in particular for networking and storage.

In this paper, we perform a detailed analysis of the performance and cost characteristics of serverless infrastructure in the data processing context. We base our analysis on a large series of micro-benchmarks across different compute and storage services, as well as end-to-end workloads. To enable our analysis, we propose the Skyrise serverless evaluation platform. For the widely used serverless infrastructure of AWS, our analysis reveals distinct boundaries for performance variability in serverless networks and storage. We further present cost break-even points for serverless compute and storage. These insights provide guidance on when and how serverless infrastructure can be used efficiently for data processing.
\end{abstract}

\maketitle

\section{Introduction}
\label{section:introduction}

Serverless infrastructure is an increasingly popular foundation for applications in the cloud \cite{cncf-report, datadog-report}. Multiple cloud providers offer compute and storage services that abstract from the provisioning and management of servers \cite{aws-serverless, azure-serverless, gcp-serverless, alibaba-serverless}. Services such as AWS Lambda \cite{lambda} and S3 \cite{s3} allocate fine-grained resources based on consumption, providing more elastic scalability and operational simplicity than conventional cloud infrastructure. The promise of cost efficiency for sporadic usage has spurred the adoption of serverless infrastructure for infrequent and short-running applications. Common examples are web, mobile, and IoT application backends that require little coordination and state management \cite{Shahrad:2020, Eismann:2022, lambda-case-studies}.

Although large-scale, data-intensive applications benefit from the elasticity of cloud infrastructure \cite{Armbrust:2009, Dageville:2016} and aim for finer-grained elasticity \cite{Melnik:2020, Vuppalapati:2020}, only few build on serverless resources in practice. We attribute this to the sub-optimal performance and cost for data access and communication, as indicated by prior work \cite{Pu:2019, Sreekanti:2020, Wawrzoniak:2021} and elaborated by this work. The elasticity of serverless architectures is enabled by storage disaggregation, which requires access to persistent and ephemeral state via the network. Today, serverless compute is offered as functions as a service (FaaS) \cite{gcf, lambda}.

Despite the limitations, data analysis systems have been built on serverless resources. The FaaS-based systems PyWren \cite{Jonas:2017, Pu:2019}, Starling \cite{Perron:2020}, and Lambada \cite{Mueller:2020} show scalable performance and cost efficiency for analytical workloads with low query volumes, i.e., inter-arrival times in minutes. For more frequent workloads, they are not cost-competitive.
To make serverless infrastructure a viable foundation for more sustained workloads, we need a better understanding of production serverless systems based on thorough evaluation. While some aspects have been studied well, e.g., CPU performance, FaaS platform overheads, and workload concurrency \cite{Scheuner:2020, Yu:2020, Copik:2021}, the factors of performance of storage and networking, and variability for processing large data require more attention.

Existing work \cite{Pu:2019, Mueller:2020, Palepu:2022} lacks a detailed analysis of the network performance of serverless functions and does not consider all serverless storage options (cf. \cite{aws-storage-options}). Most experiments are executed at small scale and it is unclear how performance translates to system components. Performance variance is a well-known issue in cloud environments \cite{Schad:2010, Uta:2020} and intensified in the serverless setting. Prior work only studies facets of serverless performance variability \cite{Perron:2020, Ustiugov:2021, Durner:2023}. We need a holistic view across application-level benchmarks, different locations, and extended timeframes. Additionally, we need a better understanding of the economic tradeoffs for using serverless cloud resources for data processing.

In this paper, we analyze the performance and cost factors of serverless infrastructure for large-scale data processing with a focus on previously overlooked characteristics. We introduce \textit{Skyrise}, an open-source framework for experimentation in serverless data processing. In summary, we make the following contributions:
\begin{enumerate}
    \item We present Skyrise, a framework to analyze the performance and cost of serverless cloud infrastructure for large-scale data processing. Skyrise includes a suite of microbenchmarks and a serverless query engine to run end-to-end workloads.\footnote{Skyrise is open-source and available at \url{https://github.com/hpides/skyrise}.}
    \item Using Skyrise, we perform an extensive evaluation of serverless networking and storage to characterize bursting and warming effects in the AWS infrastructure and show their impact on analytical applications. Factoring out these effects, we quantify remaining sources of variability.
    \item We compare the cost of query processing in Skyrise with cloud functions versus VMs and identify break-even points for the economic viability of serverless compute and storage.
\end{enumerate}

The remainder of the paper is structured as follows. In \sectionref{section:serverless_infrastructure}, we cover important background on serverless infrastructure. Then, we introduce the Skyrise framework in \sectionref{section:evaluation_framework}. In \sectionref{section:performance_evaluation}, we present our performance evaluation results. \sectionref{section:economic_viability} addresses the economic viability of serverless resources. 
We discuss our findings in \sectionref{section:discussion} and related work in \sectionref{section:related_work}. We conclude in \sectionref{section:conclusion}.  

\section{Serverless Infrastructure}
\label{section:serverless_infrastructure}

Serverless infrastructure services, such as AWS Lambda and S3, provide access to large multi-tenant pools of resources. They enable their users to consume these resources quickly, so upfront provisioning is often unnecessary to meet workload performance requirements. They bill the resources at fine granularities to minimize the cost of idle capacity. Users do not need to over-provision resources at excessive cost to prevent performance disruptions from demand spikes in dynamic workloads.

This is possible through the pervasive usage of multi-tenancy. Providers of serverless infrastructure place many user workloads on the same physical resources, e.g., CPUs or drives, to achieve high resource utilization and efficiency. They colocate uncorrelated workloads from separate applications and industries to improve the elasticity and economy of their services. Workload decorrelation enables the providers to provision for predictable cross-tenant and long-term average demand. It allows the users to employ excess capacity of co-tenants to handle demand spikes \cite{Brooker:2023}.

Sharing resources between tenants, however, leads to contention which causes variance in performance. To make their services more robust, providers use multiple techniques, including admission control, adaptive tenant placement \cite{Yanacek:2020}, and tenant isolation \cite{Bruno:2022}. While every provider implements these techniques differently, a range of serverless infrastructure services possesses inherent performance bursting, warming, and variability characteristics. Well-documented examples are:
\begin{itemize}
    \item \textbf{AWS Lambda Function Scaling:} Users can start up to 3,000 function instances in an initial burst, after which Lambda scales tenant slots at a rate of 500 per minute of load \cite{lambda-concurrency}.
    \item \textbf{AWS DynamoDB IOPS:} Users get increased burst throughput from up to 5 minutes of unused capacity. Partitions of tenants that constantly exceed capacity are migrated \cite{Elhemali:2022, dynamodb-practices}.
    \item \textbf{Google Cloud Storage:} Users need to gradually increase request rates to warm up their buckets for load spikes \cite{gcp-storage-practices}. They must expect high tail latencies  for requests \cite{Ustiugov:2021}.
\end{itemize}

The rest of this section presents the two types of infrastructure that are most essential for data processing, namely FaaS platforms providing compute capacity and serverless storage.

\subsection{Function as a Service Platforms}
All major cloud vendors have FaaS platforms, such as AWS Lambda, Google Cloud Functions \cite{gcf}, Azure Functions \cite{maf}, and Alibaba Cloud Function Compute \cite{alibaba-fc}. Users of cloud function services upload their application binaries as ZIP archives or container images. They configure function sizes and how functions are invoked.

\begin{table}
    \centering
    \caption[Configuration and pricing of AWS compute services.]{Configuration and pricing of AWS compute services.}
	\vspace{-0.3cm}
    \setlength\tabcolsep{5.3pt}
    \renewcommand{\arraystretch}{0.8}
    \begin{tabular}{lrr}
      \toprule
      \textbf{Resource} & \textbf{Lambda} (ARM) & \textbf{EC2} (C6g\tablefootnote{ARM-based Lambda functions \cite{lambda-isas} and compute-optimized EC2 C6g instances \cite{ec2-c6g} use Graviton2 processors \cite{graviton2} and have comparable CPU-to-RAM capacity ratios.}) \\
      \midrule
        \textbf{Memory} & Configurable \cite{lambda-limits} & Configurable\tablefootnote{The resource capacity ratios and limits depend on the EC2 instance type \cite{ec2-instance-types}.} \\
        Capacity [GiB] & 0.125 -- 10 & 2 -- 128 \\
        Price [\textcent/GiB-h]\tablefootnote{We provide ranges for the pricing tiers of Lambda \cite{lambda-pricing} and the prices of EC2 on-demand and reserved instances \cite{ec2-pricing}. All prices apply to the AWS us-east-1 region.} & 3.84 -- 4.80 & 0.65 -- 1.70 \\
      \midrule
        \textbf{Compute} & Memory-based\tablefootnote{A Lambda function gets 1 vCPU equivalent per 1,769 MiB of memory \cite{lambda-limits, lambda-isolation}.} & Configurable \\
        Capacity [vCPU]
        & 0.07 --  5.79 & 1 -- 64 \\
        Price [\textcent/vCPU-h] & 6.79 -- 8.49 & 1.30 -- 3.40 \\
      \midrule
        \textbf{Network} & Constant & Compute-based \cite{ec2-network} \\
        Bandwidth [Gbps]\tablefootnote{AWS only provides partial information \cite{lambda-limits,ec2-c6g}. We report the baseline bandwidth from \sectionref{section:network_bursting}. Note that Lambda network bandwidth is constant over instance sizes.} & 0.63 & 0.375 -- 25 \\
        Price [\textcent/Gbps-h] & 0.48 -- 0.60 & 3.27 -- 8.70 \\
      \midrule
        \textbf{Storage} & Configurable & Configurable \\
        Capacity [GiB] & 0.5 -- 10 & 0 -- 3,800 \\
        Price [\textcent/GiB-mo] & 8.12 & 2.33 -- 5.41 \\
      \bottomrule
    \end{tabular}
    \label{table:2_compute_services}
    \vspace{-0.2cm}
\end{table}

Current FaaS platforms limit their configuration, as is shown in \tableref{table:2_compute_services}. Function sizing is usually done based on memory capacity, which determines the number of virtual CPUs. Compared to VMs, functions are restricted to be about an order of magnitude smaller. FaaS platforms further disallow hour-long lifetime, persistent state, and direct communication. Functions with these characteristics are small, short, and ephemeral tasks that are easy to manage.

\begin{figure}[bt]
    \centering
    \includegraphics[width=0.9\linewidth]{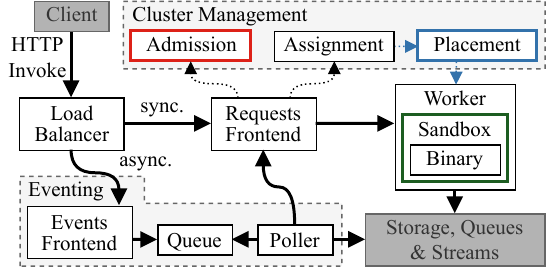}
    \vspace{-0.4cm}
    \caption{Architecture of the AWS Lambda FaaS platform with control (dotted) and data (solid) paths.}
    \vspace{-0.45cm}
    \label{figure:lambda_architecture}
\end{figure}

Cloud functions are invoked by either users via HTTP requests or triggers on events from queues \cite{sqs, kafka}, streams \cite{kinesis, Carbone:2015}, and storage services. The architecture and invocation procedure of the most widely used FaaS system Lambda is depicted in \figureref{figure:lambda_architecture} \cite{Agache:2020}. A user request enters the system via a load-balanced frontend service, which coordinates function invocations. The frontend retrieves the function metadata and checks with the admission service (in \textcolor{red}{red}) if the invocation exceeds the user's quota for concurrent function executions. Then, the frontend asks the assignment service for a sandboxed environment in the worker fleet to execute the function binary. Sandboxes (in \textcolor{green}{green}) are implemented with virtualization  to run thousands of functions from different users in isolation on each worker machine \cite{lambda-isolation, Agache:2020}. If there is a free sandbox, the frontend routes the request payload to this sandbox for function execution. If no sandbox is available, the placement service is asked to create a new environment on a worker with sufficient capacity. This involves downloading and initializing the binary along with its dependencies and is referred to as a coldstart (the \textcolor{blue}{blue} path). Other than synchronous requests, both asynchronous requests and events are received by the polling service which polls their payloads from internal and external queues, respectively, and invokes functions as a proxy, adding further latency to the invocation path.

FaaS platforms charge based on the number, size, and duration of function invocations \cite{lambda-pricing}. The duration is typically tracked at a millisecond granularity, as opposed to the second to minute intervals in VM billing \cite{ec2-pricing}. Depending on the resource, Lambda functions have 2.5--5.9$\times$ higher unit prices than EC2 VMs (cf. \tableref{table:2_compute_services}).

\subsection{Serverless Storage Services}
\label{section:serverless_storage}

Cloud users have three options for serverless storage: object stores \cite{s3, mabs}, key-value stores \cite{ddb-serverless, gcp-firestore}, and shared filesystems \cite{efs-serverless, mafiles}. All of these services provide elastically scalable storage capacity with high availability and durability guarantees.

Object storage, such as S3 and Azure Blob Storage, is designed to store immutable binary objects of varying sizes and to access these objects with scalable bandwidth. Key-value stores, such as DynamoDB and GCP Firestore, support lower latency key lookups at higher IOPS for kilobyte-sized values. AWS EFS, Azure Files, and other networked filesystems with an NFS or SMB interface provide an abstraction of files and directories.

\begin{figure}[bt]
    \centering
    \includegraphics[width=0.7\linewidth]{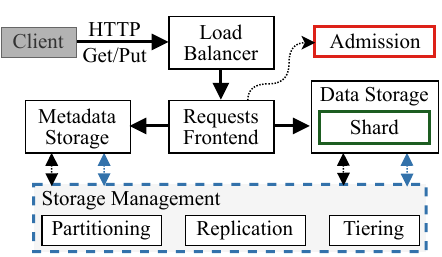}
    \vspace{-0.4cm}
    \caption{Schematic architecture of a serverless storage system with control (dotted) and data (solid) planes.}
    \vspace{-0.6cm}
    \label{figure:storage_architecture}
\end{figure}

The high-level architecture of serverless key-value and object storage systems is illustrated in \figureref{figure:storage_architecture} \cite{Elhemali:2022, Warfield:2023}. Users interact with these systems via a simple HTTP Get/Put API. Their requests go through load-balancing, admission (marked \textcolor{red}{red}), and request-routing components before a metadata service maps the requested key to a server in the storage fleet. Then, the storage server receives or sends the data. Data is partitioned and replicated across many storage servers and servers hold shards (in \textcolor{green}{green}) of many users. Partitions that outgrow their size or serve excessive load are split and spread evenly across the fleet (marked in \textcolor{blue}{blue}). We refer to this process as warming and the inverse process as cooling. For instance, S3 partitions serve 3.5--5.5K IOPS before being split \cite{s3-practices}.

\begin{table}[bt]
    \centering
    \caption[Pricing of AWS serverless storage services.]{Pricing of AWS serverless storage services.}
    \vspace{-0.3cm}
    \setlength\tabcolsep{5.8pt}
    \renewcommand{\arraystretch}{0.8}
    \begin{tabular}{lrrrrr}
      \toprule
      \textbf{Component} & \multicolumn{2}{r}{\textbf{Requests}} & \multicolumn{2}{r}{\textbf{Transfers}} & \textbf{Storage} \\
      & \multicolumn{2}{r}{[\textcent/M]} & \multicolumn{2}{r}{[\textcent/GiB]\tablefootnote{We assume no costs for service access across regions, zones, or virtual networks \cite{aws-transfers}.}} & [\textcent/GiB-mo]\tablefootnote{We provide ranges for the pricing tiers of S3 \cite{s3-pricing} and the prices of EFS types \cite{efs-pricing}.} \\
      & Read & Write & Read & Write & \\
      \midrule
        S3 Standard & 40 & 500 & 0 & 0 & 2.1 -- 2.3 \\
        S3 Express\tablefootnote{The S3 Express storage class charges requests for transferred data beyond 512 KiB.} & 20 & 250 & 0.15 & 0.8 & 16 \\
        DynamoDB & 25 & 125 & 0 & 0 & 25 \\
        EFS & 0 & 0 & 3 & 6 & 16 -- 30 \\
      \bottomrule
    \end{tabular}
    \label{table:2_storage_services}
    \vspace{-0.3cm}
\end{table}

The pricing model of serverless storage services is a composite of prices for data storage, requests, and transfers, as shown in \tableref{table:2_storage_services}. In AWS, S3 is by an order of magnitude the cheapest option to store data. S3 request cost are independent of the size (from 1 B to 5 TiB), yet they are the highest among the services. Keeping S3 warm for 100K IOPS costs \$144 per hour. The pre-warmed S3 Express variant charges requests based on size, resulting in 24--115$\times$ higher prices for the throughput-optimal 8--16 MiB range \cite{s3-practices}. In DynamoDB, requests are split and charged in kilobyte-scale units. EFS has the highest data transfer fee.

\section{Skyrise Evaluation Framework}
\label{section:evaluation_framework}

In this section, we introduce the Skyrise evaluation framework for experimentation in serverless data processing. Our framework includes a comprehensive suite of microbenchmarks for serverless resources and integrates a serverless query execution engine to run application-level benchmarks. The framework automates the setup, execution, and result processing for the experiments in our evaluation. Hence, it enables the reproduction of our experimental results. We give an overview of the design and implementation of the framework in \sectionref{section:framework_overview}. Then, we describe our prototypical query engine for the execution of full queries in \sectionref{section:query_engine}.

\subsection{Framework Overview}
\label{section:framework_overview}

\begin{figure}[bt]
    \centering
    \includegraphics[width=0.8\linewidth]{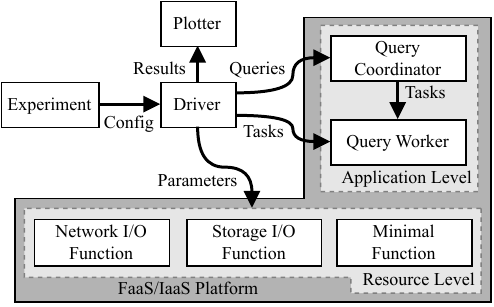}
    \vspace{-0.2cm}
    \caption{Architecture of the Skyrise evaluation framework showing the experiment execution flow from config to plot.}
    \vspace{-0.6cm}
\label{figure:skyrise_evaluation_framework}
\end{figure}

The core components of our framework (cf. \figureref{table:3_experiment_overview}) are written in C++ and Python. We use C++ for the performance-critical system drivers and measurement functions. The experiment configurations and result plots are in Python. We integrate our codebase with the AWS infrastructure services Lambda and EC2 for compute, as well as S3, DynamoDB, and EFS as storage options. We choose the AWS environment, because Lambda is the most widely used \cite{datadog-report} and studied \cite{Scheuner:2020} FaaS platform with the fewest performance-related restrictions \cite{lambda-limits, maf-limits, gcf-limits}. The framework is open-source and enables the integration of additional benchmarks and cloud infrastructure.

The framework supports experiments on two levels in the stack. On the resource level, the framework employs microbenchmarks measuring performance metrics of compute and storage services. For the application level, e.g., query operators or complete queries, the framework uses our query engine.

\begin{table*}[bt]
    \centering
    \caption[Overview of experiment configurations.]{Overview of experiment configurations.}
    \vspace{-0.3cm}
    \setlength\tabcolsep{7.5pt}
    \renewcommand{\arraystretch}{0.8}
    \begin{tabular}{lllll}
      \toprule
      \textbf{System under Test} & \textbf{Driver} & \textbf{Functions} & \textbf{Parameters} & \textbf{Metrics} \\
      \midrule
      Lambda & FaaS Platform & Minimal, Network I/O, & Instance Size \& Count & I/O Throughput, Startup Latency,\\
      & & Storage I/O & & Idle Lifetime \\
      EC2 & IaaS Platform & Network I/O, Storage I/O & Instance Type \& Count & I/O Throughput, Startup Latency \\
      S3, DynamoDB, EFS & IaaS \& FaaS & Storage I/O & File Size \& Count & I/O Throughput, IOPS, Latency  \\
      \midrule
      Skyrise Query Engine & Data System & Query Coordinator, & Queries, Data Size, & Query Latency \& Cost \\
      & & Query Worker & Deployment &  \\
      \bottomrule
    \end{tabular}
    \label{table:3_experiment_overview}
    \vspace{-0.3cm}
\end{table*}

To execute experiments, the framework deploys and invokes cloud function binaries. Figure~\ref{figure:skyrise_evaluation_framework} gives an overview of this process. Each experiment defines a configuration, which is submitted to a driver. Depending on the experiment level, the driver invokes a specific function binary on its target platform. For resource-level microbenchmarks, the driver invokes one or more instances of the following cloud functions.

\labeltitle{Network I/O.} To analyze network performance in isolation, the network measurement function uses iPerf3 \cite{iperf3}, an open-source network performance measurement tool. Our function employs the C API, which allows fine-grained parameter tuning. The function sends or receives randomly generated data for a pre-specified time.
\labeltitle{Storage I/O.} The storage I/O measurement function writes or reads randomly generated files of fixed size and number to or from a storage service. For latency measurements, the function calls the synchronous storage service APIs. For throughput measurements, it calls the asynchronous APIs from a fixed-size thread-pool.
\labeltitle{Minimal.} This binary incorporates the minimum amount of code for a cloud function and is a no-op. It does not link any libraries, but random BLOBs of pre-specified sizes for startup experiments.

For application-level experiments, the driver runs queries with our query engine by calling the query coordinator function, which in turn breaks queries into tasks and schedules worker functions for them. We employ the following suite of queries.

\labeltitle{Queries.} Our query suite includes TPC-H Q1, Q6, and Q12, as well as TPCx-BB Q3. These queries are I/O-heavy and thus lend themselves well to evaluate cloud resources. We specifically avoid queries that benefit from sophisticated optimization or execution techniques that hide resource aspects. Q1 and Q6 select, project, and aggregate data. Q3 and Q12 are join queries with a broad set of operators, including user-defined functions (UDFs).

\tableref{table:3_experiment_overview} gives an overview of the experiment configurations. For configurations targeting EC2, we use a shim layer that resembles the Lambda execution environment to run functions on VM hosts.

When the experiment ends, the driver receives result metrics from multiple sources: Logs, traces, and a response from the invoked function. For resource-level experiments, the metrics include, e.g., timestamps, request counts, latencies, and throughputs. The driver then aggregates these results and estimates the experiment cost using the AWS price list service, disregarding any bulk discounts. For application-level experiments, the query coordinator function returns high-level metrics such as query latency and cost. Finally, the driver stores the results in a JSON file and hands them to a plotter for visualization.

\subsection{Query Execution Support}
\label{section:query_engine}

\begin{figure}[bt]
    \centering
    \includegraphics[width=0.9\linewidth]{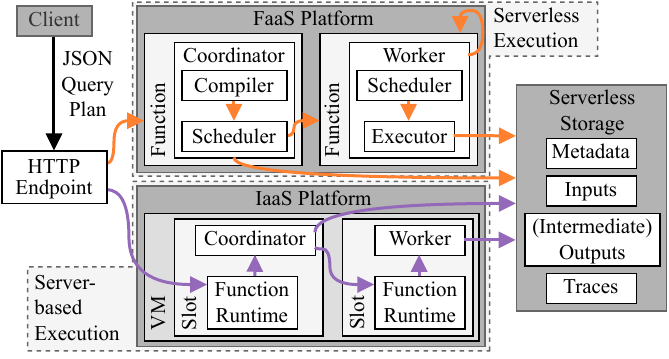}
    \vspace{-0.3cm}
    \caption{Execution modes of the Skyrise query engine with serverless functions (\textcolor{orange}{upper path}) and servers (\textcolor{purple}{lower path}).}
    \vspace{-0.75cm}
    \label{figure:skyrise_query_engine}
\end{figure}

The evaluation framework integrates a serverless query engine \cite{Bodner:2020} to support the execution of end-to-end workloads. This allows to understand how resource level effects translate to application performance. The query engine is designed to run entirely on serverless infrastructure. Figure \ref{figure:skyrise_query_engine} shows an overview of its shared-storage architecture with state kept in serverless storage. The coordinator and worker nodes are deployed as serverless functions and employ shared serverless storage to load inputs and communicate outputs. As a deployment alternative, the query engine can run on VMs, enabling the direct comparison of FaaS and IaaS-based execution.

To execute a query, the framework's driver sends a physical query plan in JSON format to an HTTP endpoint \cite{lambda-urls}. On an FaaS platform, this triggers a serverless function running the coordinator. In an IaaS deployment, the request is routed to the same coordinator binary yet running on a provisioned VM with our shim layer. In both cases, the query plan is passed to the coordinator. A plan contains pipelines of physical operators as well as the dependencies between the pipelines. The coordinator fetches the metadata on the referenced pipeline input datasets, including the number and sizes of the files. The coordinator then compiles a distributed query plan, deciding on the number of fragments per pipeline for data-parallel execution and on worker sizing. This plan is the same for FaaS and IaaS deployments. Next, the coordinator schedules the pipelines stage-wise based on their dependencies. In serverless execution mode, the scheduler invokes a worker function for each pipeline fragment. In server-based execution, it queues and distributes the fragments across the available worker slots. A worker parses its query fragment and schedules the operators for execution. Workers use a vectorized execution model. The execution includes reading input data partitions in batches from shared storage, generating partitioned outputs, and writing them back to storage. Upon completion of the final query pipeline, the coordinator returns a JSON response with the location of the query result in serverless storage, the query runtime and cost. This response is wrapped and sent back to the HTTP client. 

To isolate and analyze query subflows, such as distributed scans and shuffles, the engine supports the injection of synchronization barriers into its execution. This mechanism is implemented as an extra operator that polls a shared queue for a barrier condition.

Moreover, the engine traces runtime information with query context. This information can be compared between distributed workers, as their clocks are tightly synchronized \cite{Bodner:2022, aws-clocks}.

Finally, the query engine adopts a number of techniques for an execution performance that is representative of other cloud-based data processing systems. They fall broadly into two categories.

\labeltitle{Exploiting Serverless Compute Elasticity.} To start up a large cluster of workers quickly, the engine employs a two-level function invocation procedure \cite{Mueller:2020}. Scheduling 256 or more workers, the coordinator parallelizes function calls across a subset of workers.

To reduce function invocation latency, we decrease the duration of coldstarts and increase the probability of warmstarts. We keep binary sizes small (< 10 MiB) by linking against the library versions present in Lambda sandboxes and stripping off any unneeded symbols. The deployment artifacts are not specialized towards any query. As such, they can be reused as long as they are cached in the FaaS platform sandboxes. 

\labeltitle{Efficient Data Access on Serverless Storage.} To execute queries efficiently over data stored in columnar file formats on cloud storage, the engine divides large storage requests into smaller chunks to process them in parallel. Straggling requests are retriggered after a size-based timeout. Parquet \cite{parquet} and ORC \cite{orc} file metadata is read to identify relevant data and push down projections and selections.

\section{Performance Evaluation}
\label{section:performance_evaluation}

In this section, we present the results of our experiments using our evaluation framework. We give insights into both low-level and workload-specific performance of serverless cloud infrastructure to better understand its potential for data processing. Specifically, we examine the following aspects in the corresponding sections:
\begin{itemize}
    \item The dominant source of performance variability in serverless function networking (\sectionref{section:network_bursting})
    \item The choice of storage for serverless analytics and the major source of variable performance (\sectionsref{section:storage_comparison}{section:iops_scaling})
    \item The control and translation of above aspects to application-level performance (\sectionref{section:exploiting_characteristics})
    \item A quantification of the remaining variability (\sectionref{section:performance_predictability})
\end{itemize}

\subsection{Experimental Setup}
\label{section:setup}
We conduct all experiments on AWS in the availability zone (AZ) us-east-1a \cite{aws-zones} in a single virtual private cloud (VPC \cite{aws-vpc}) network  and in the time frame of February to October 2024 unless stated otherwise. For our experiments, we deploy ARM-based Lambda functions and on-demand VMs of the EC2 C6g family of instances\footnote{C6g instances are still much more widely deployed than the more recent C7g family \cite{ec2-c7g, ec2-availabilty} and thus easier to provision in large numbers (>100) for our experiments.}. Our driver has low resource requirements and runs continuously on a c6g.xlarge instance across experiments. All other compute resources are newly created for each experiment configuration and repetition. To run our large-scale experiments, we asked AWS to increase our account quotas for parallel function invocations as well as vCPUs in our VM fleet to 10,000 and 5,000 \cite{lambda-limits, aws-quotas}. As serverless storage services, we consider S3, S3 Express \cite{s3-eoz}, DynamoDB, and EFS. We do not consider the deprecated S3 Select service \cite{s3-select}. In our evaluation, we do not include VM-based container services (like the Elastic Kubernetes Service \cite{eks}) or provisioned storage options (like ElastiCache \cite{elasticache}). We also do not include the serverless offerings from other cloud providers due to budget constraints.

We track service usage via a client hook that counts all requests, including failures and retries. Based on the request counts and the runtimes of the employed cloud functions and VMs, we calculate and report the experiment cost. In total, our experiments perform millions of function invocations and billions of storage requests moving hundreds of terabytes of data, adding up to around \$4,000.

\subsection{Burstable Function Network Bandwidth}
\label{section:network_bursting}

Distributed data systems require high network throughput. This is particularly the case for systems with stateless compute and disaggregated storage, as they repeatedly scan large amounts of data from remote storage. In addition, serverless query engines usually shuffle intermediate results between workers via remote storage. Consequently, both scanning and shuffling rely on the network to transfer the data. However, the network performance of cloud functions is opaque. Unlike VMs, cloud vendors do not expose the network bandwidth that functions can achieve.

In this experiment, we employ network I/O functions as clients while we deploy iPerf3 servers on EC2 instances with increased network throughput \cite{ec2-c6g, ec2-network}, so that they do not become the bottleneck. A single server serves up to 10 clients. A well-known limitation of EC2 is the 5 Gbps limit for single-flow network traffic. To bypass this limitation and explore the full potential bandwidth of functions, we establish multiple paths, i.e., TCP connections, between each pair of endpoints. We utilize functions with 4 vCPUs, and allocate one TCP connection per vCPU. This allows us to detect a theoretical network bandwidth of up to 20 Gbps per function \cite{ec2-network}.

\subsubsection{Network Bursting}
\begin{figure}[bt]
    \centering
    \includegraphics[width=1.0\linewidth]{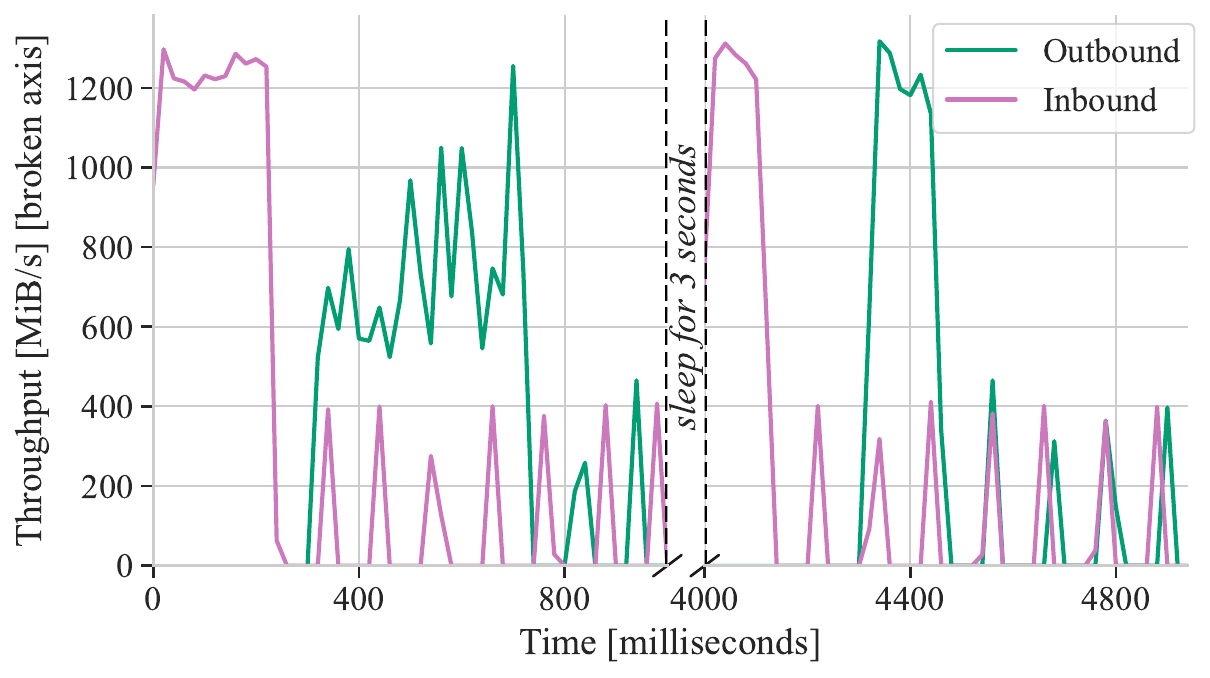}
    \vspace{-0.6cm}
    \caption{Function network throughput at 20 ms intervals with short sleep to refill inbound/outbound token buckets.}
    \vspace{-0.5cm}
    \label{figure:lambda-network-token-bucket}
\end{figure}

We run the network microbenchmark for five seconds, including an intermittent break of three seconds with no traffic. We repeat the experiment ten times and plot the run with the median network throughput. \figureref{figure:lambda-network-token-bucket} reveals that a function initially delivers an inbound bandwidth of 1.2 GiB/s and is capable of maintaining this bandwidth for 250 milliseconds. Afterwards, the bandwidth drops to zero and shows regular spikes in which data is transferred. We observe that this burst is renewable, i.e., it reoccurs after the 3-second break, yet the duration of the second burst is shorter. A similar picture emerges for the outbound bandwidth, although the bandwidth is reduced and shows higher variation. We attribute this to the additional overhead of data generation of the iPerf3 library. We conclude that the inbound and outbound token buckets are maintained independently of each other.

This is consistent with the rate limiting parameters of the virtual machine monitor that Lambda builds on, and according to our throughput measurements, both buckets are configured with an initial capacity of {\raise.17ex\hbox{$\scriptstyle\mathtt{\sim}$}}300 MiB.
Once the token bucket empties, 7.5 MiB of data can be consumed in 100 millisecond intervals, resulting in a baseline bandwidth of 75 MiB/s. Furthermore, we find that the token bucket refills halfway to the initial capacity as soon as a function stops utilizing the network or terminates. This implies that a one-off, non-rechargeable budget of {\raise.17ex\hbox{$\scriptstyle\mathtt{\sim}$}}150 MiB exists in addition to a rechargable token bucket capacity of {\raise.17ex\hbox{$\scriptstyle\mathtt{\sim}$}}150 MiB.

\begin{figure}[bt]
    \centering
    \includegraphics[width=1.0\linewidth]{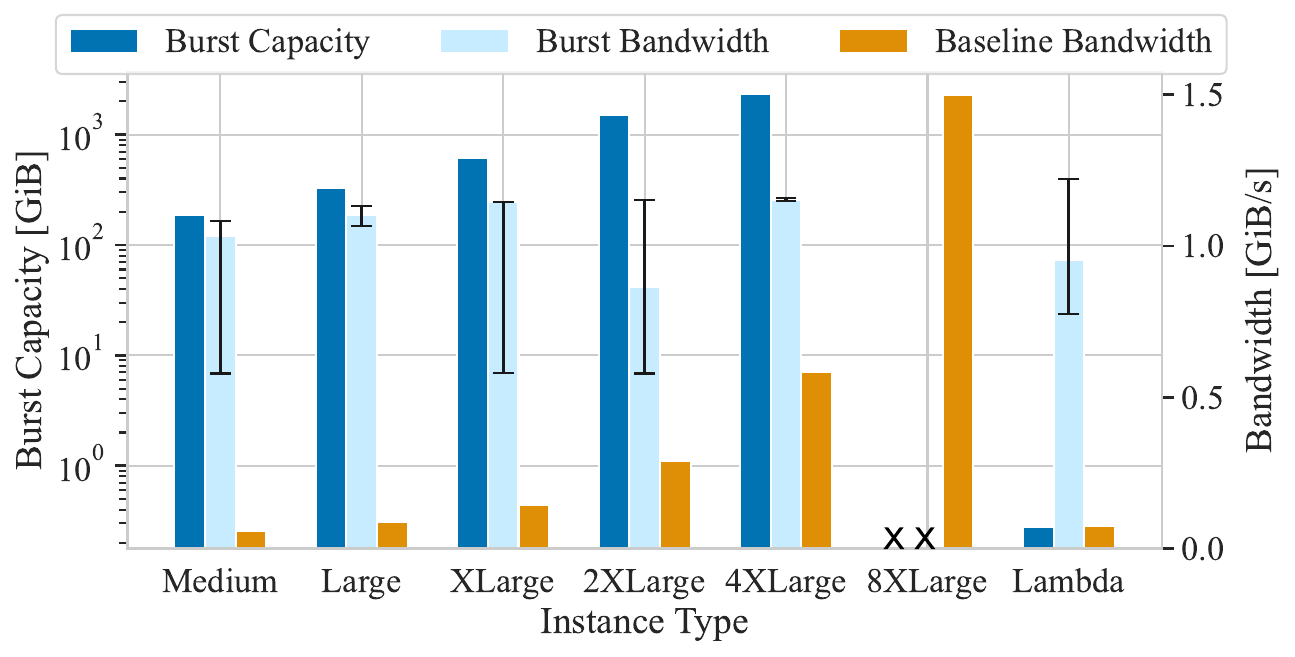}
    \vspace{-0.6cm}
    \caption{EC2 C6g and Lambda network bursting behavior with burst and baseline throughput, and token bucket size.}
    \vspace{-0.5cm}
    \label{figure:lambda-ec2-network-bursting}
\end{figure}

We rerun our microbenchmark with EC2 VMs of varying sizes as clients. The benchmark duration depends on the VM size and ranges from three to 45 minutes. We run the experiment three times per configuration and again report the median burst throughput. \figureref{figure:lambda-ec2-network-bursting} shows how the network performance of Lambda and EC2 compares. We report the initial capacity of the token bucket for the burst mechanism in GiB. We also report the bandwidth under burst and the sustained baseline bandwidth. Both services employ a bursting mechanism. Lambda allows throughput to burst for a short period of time, while the token bucket size of EC2 instances and the duration of their burst are substantially longer and increases with instance size. We see a high variation for both EC2 and Lambda network burst throughputs, yet very stable burst capacities.

\subsubsection{Scalable Network Performance}

\begin{figure}[bt]
    \centering
    \includegraphics[width=1.0\linewidth]{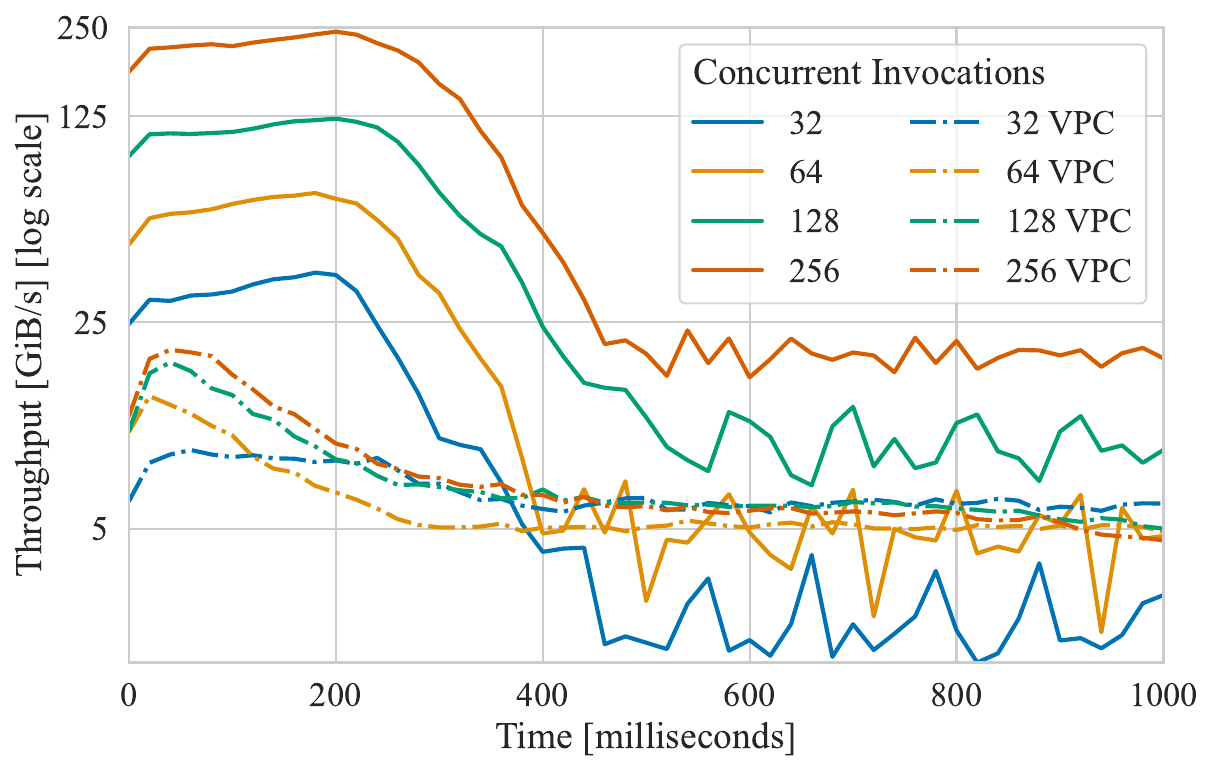}
    \vspace{-0.6cm}
    \caption{Aggregated network throughput at 20 ms intervals for varying concurrency (32 to 256) and with/without VPC.}
    \vspace{-0.5cm}
    \label{figure:lambda-network-scale-out}
\end{figure}

To study the scaling behavior of Lambda's bursting network performance, we conduct another experiment mapping 32 to 256 functions on a cluster of (4--26) iPerf3 servers. We measure the aggregated network throughput in two different settings. As organizations often deploy their applications in customer-owned VPCs, we restrict parts of the experiment to a VPC within a single AZ. We omit the customer-owned VPC in the second part of the experiment and compare the results of both settings. \figureref{figure:lambda-network-scale-out} shows that the baseline and burst bandwidths scale horizontally. We attribute this to the ability of Lambda to place functions effectively. However, we observe limited scalability if the experiment runs in a customer-owned VPC within a single AZ. In particular, we see a hard limit of {\raise.17ex\hbox{$\scriptstyle\mathtt{\sim}$}}20 GiB/s in throughput. Restricting the service to deploy functions within a VPC appears to either hinder its scheduling flexibility or to introduce a network throughput quota. We conclude that the burstable network throughput of Lambda is significant, deterministic, and scalable (outside of VPCs). Data processing systems should thus aim to exploit it.

\subsection{Comparing Serverless Storage Options}
\label{section:storage_comparison}

Object stores, key-value stores, and shared filesystems are seen to be complementary in their performance characteristics in terms of throughput, IOPS, and latency. To satisfy diverse I/O requirements, data-intensive applications often build on a combination of these options \cite{Armenatzoglou:2022, aws-dropbox}. However, there is little empirical research on when to use which option over another \cite{Roy:2021, Palepu:2022} or when none of them are sufficient, leaving a gap in the serverless storage landscape \cite{Klimovic:2018, Pu:2019}. For this reason, we conduct a comprehensive comparison of the current serverless storage options on AWS. We study the performance and price tradeoffs between the S3 object store, the DynamoDB key-value store, and the network filesystem EFS in the context of large-scale data processing. We evaluate these storage services for throughput of up to hundreds of gigabytes and hundreds of thousands of requests per second. In addition, we analyze their latency distribution over millions of requests.

For these experiments, we employ our storage I/O function. The function runs on EC2 VMs, because EC2 burst capacity allows for sustained high bandwidth throughout the experiments. We use c6gn.2xlarge instances with eight vCPUs, 16 GiB memory, and burstable network bandwidth of up to 25 Gbps. In experiments with distribution, all instances synchronize via a shared queue upon startup to ensure concurrent execution. To reduce storage-side scaling and caching effects, we keep repetition durations short (<5 minutes) and intervals between repetitions long (>12 hours). Unless otherwise noted, we present the median out of three repetitions.
In our comparison, we consider the S3 Standard and Express storage classes. We further include DynamoDB with on-demand capacity \cite{dynamodb-capacity} and strongly consistent reads \cite{dynamodb-consistency}, as well as EFS with elastic throughput and synchronous writes \cite{efs-performance}. All services only charge for consumption and provide at least the read-after-write consistency model of S3 \cite{s3-consistency, Kingsbury:2024}.

\subsubsection{Throughput}
\label{section:storage_throughput}
\begin{figure}[bt]
    \centering
    \includegraphics[width=1.0\linewidth]{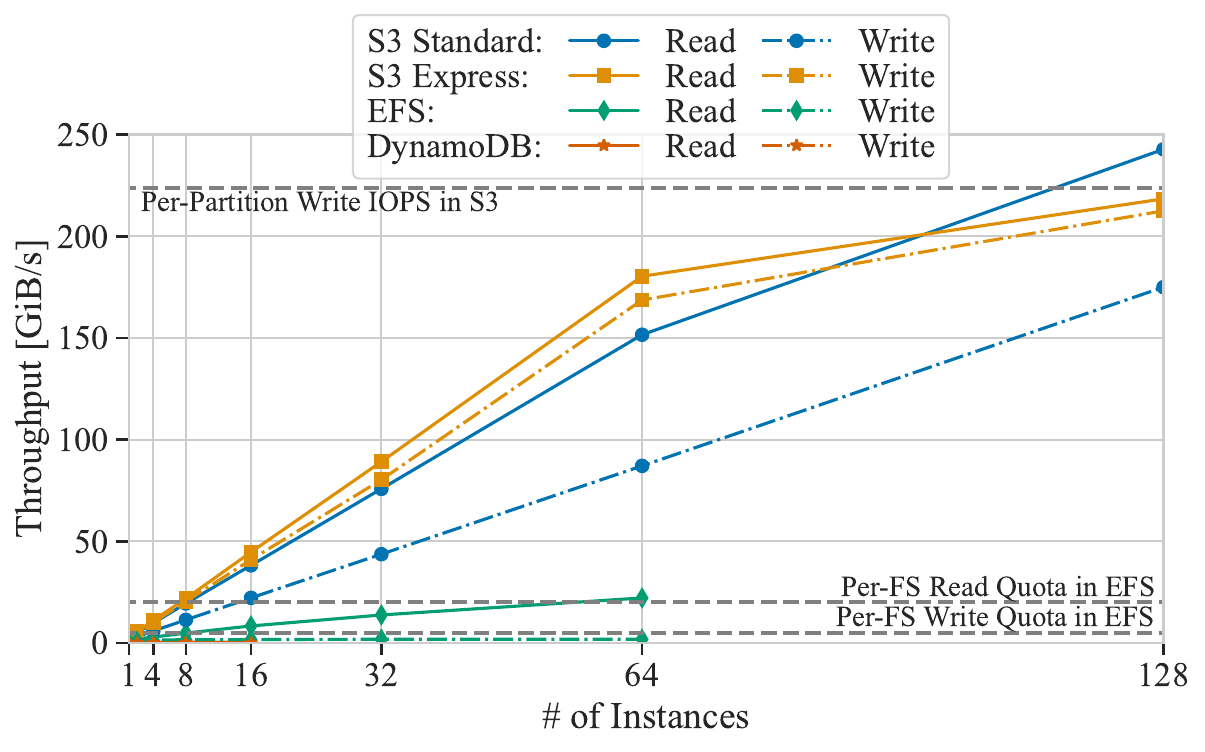}
    \vspace{-0.8cm}
    \caption{Aggregated read/write throughput of the serverless storage services for varying number of client  VMs (1 to 128).}
    \vspace{-0.6cm}
    \label{figure:storage_throughput}
\end{figure}

Serverless systems scan and shuffle data through storage. Both scanning and shuffling are throughput-heavy operations. In this experiment, we investigate the scalability of throughput of the storage services with the number of compute nodes generating load. We schedule up to 128 nodes running 32 I/O threads. For S3, we generate and access 64 MiB objects, allowing for roughly 250 GiB/s of throughput on an unpartitioned bucket \cite{s3-practices}. For DynamoDB, we employ the largest possible item size of 400 KiB. We access a single table, since sharding over multiple new on-demand tables does not yield higher throughput. We write 4 MiB files to EFS and read them back with no caching.

Our results in \figureref{figure:storage_throughput} show that both the S3 variants scale linearly up to the generated load of {\raise.17ex\hbox{$\scriptstyle\mathtt{\sim}$}}250 GiB/s. We attribute the difference in their write throughput to the less consistent IOPS performance of standard S3. The serverless versions of DynamoDB and EFS fall short of the target throughput. They each start rejecting requests under contention at different degrees of concurrency. EFS serves up to 64 client VMs with its throughput converging to the quotas (20 and 5 GiB/s \cite{efs-performance}) for an individual filesystem instance. DynamoDB's throughput is already saturated by a single client VM and stays at {\raise.17ex\hbox{$\scriptstyle\mathtt{\sim}$}}380 MiB/s for reading and {\raise.17ex\hbox{$\scriptstyle\mathtt{\sim}$}}30 MiB/s for writing until most requests get throttled or time out at around 16 clients. 

Taking price (cf., \tableref{table:2_storage_services}) into account, S3, DynamoDB, and EFS cost 0.00064, 6.55, and 3.00 \textcent/GiB/s for reading data, respectively. This makes S3 also the by far most cost-efficient option.

\subsubsection{Operations per Second}
\label{section:storage_iops}

\begin{figure}[bt]
    \centering
    \includegraphics[width=1.0\linewidth]{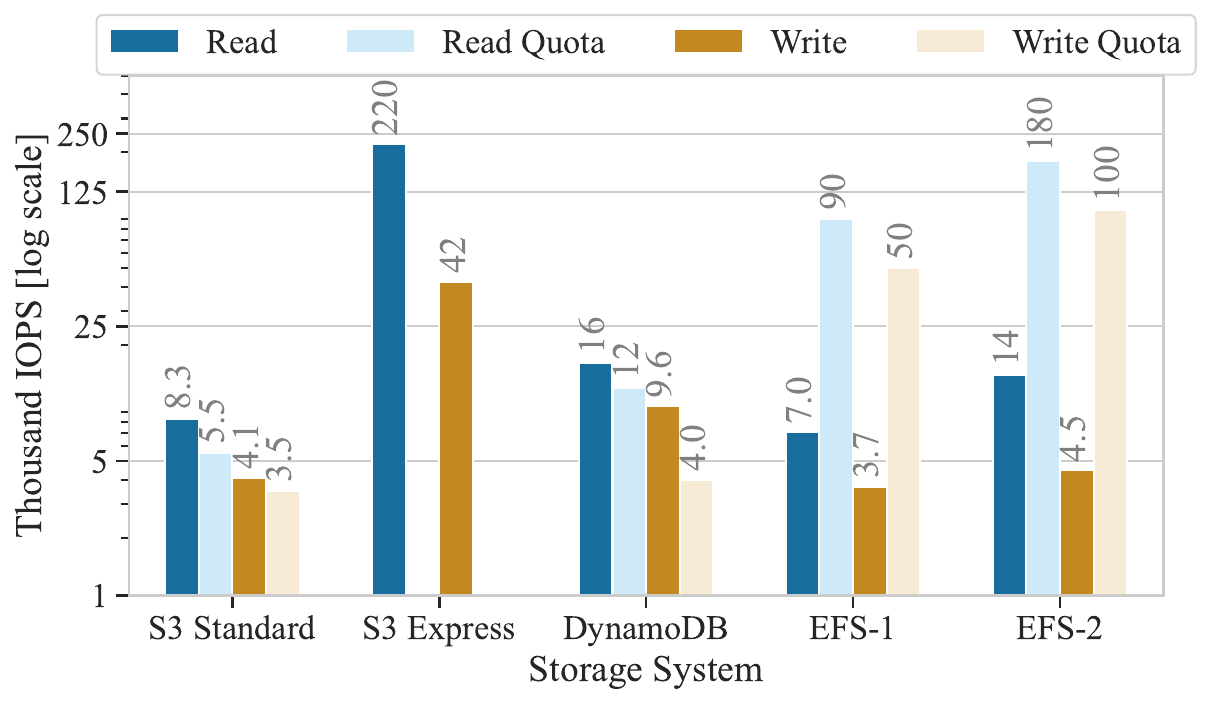}
    \vspace{-0.8cm}
    \caption{Operations per second and container-level quotas for each serverless storage system. For EFS, we present configurations with one (EFS-1) and two (EFS-2) filesystems.}
    \vspace{-0.5cm}
    \label{figure:storage_iops}
\end{figure}

To process queries on large datasets, cloud analytics systems need to access many individual objects or files. This requires to send large numbers of concurrent requests to the storage services. We measure the IOPS performance of S3, DynamoDB, and EFS on newly created buckets, tables, and filesystems. We again run up to 128 nodes, each with 32 dedicated threads sending 1 KiB requests for a total of >250K requests per second. For DynamoDB, we asked AWS to increase the table and account-level IOPS quotas to 250K.
Our results are shown in \figureref{figure:storage_iops}. The standard S3 performance is just above the target IOPS for an individual prefix partition \cite{s3-practices} with 8K reads and 4K writes per second. S3 Express is not subject to the partition quota, providing the highest IOPS in our comparison with 220K for reads and 42K for writes. DynamoDB also provides slightly more IOPS than defined by the quotas for new on-demand tables \cite{dynamodb-capacity}, with 16K read IOPS and 9.6K write IOPS. We miss the per-filesystem quotas of EFS by more than an order of magnitude, despite closely following the documentation \cite{efs-performance}. The read IOPS double via sharding over two filesystems, but do not scale further.

\subsubsection{Latency}
\begin{figure}[bt]
    \centering
    \includegraphics[width=1.0\linewidth]{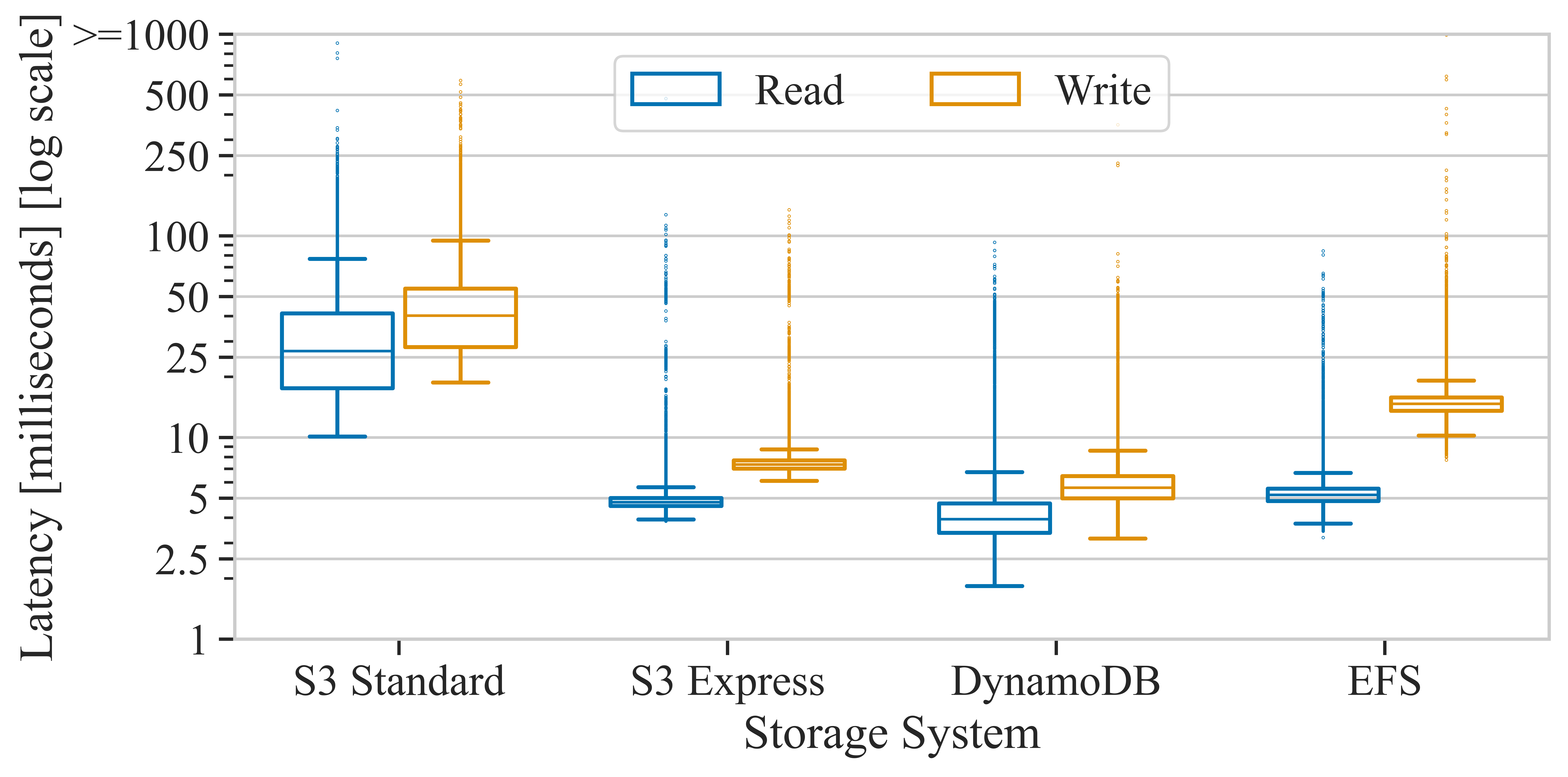}
    \vspace{-0.8cm}
    \caption{Latency distribution of each serverless storage system for one million read/write requests including outliers.}
    \vspace{-0.5cm}
    \label{figure:storage_latency}
\end{figure}

Compared to compute local storage, disaggregated cloud storage entails orders of magnitude higher latency. While analytics are usually throughput-bound, latency is still important and compensated via increased request size or concurrency \cite{Mueller:2020,Durner:2023}. To determine the latency distributions of S3, DynamoDB, and EFS, we send one million 1 KiB read and write requests to every service. To keep the experiment duration moderate and the load on the services low, we employ 10 clients using the synchronous APIs \cite{aws-sdk-cpp}. We present our results in \figureref{figure:storage_latency}. We observe that S3 Standard has both the highest median (27 ms for reads and 40 ms for writes) and tail latencies. Out of 1M read requests, 95\% completed in 75 ms and the slowest requests took just over 10 s (374X of the median). S3 Express benefits from its zonal deployment \cite{s3-eoz} and provides significantly lower and less variable latencies with the median and 95th percentile read latencies at around 5 ms. DynamoDB exhibits slightly lower yet more variable latencies than S3 Express. Finally, EFS provides similarly low and consistent read request latencies as S3 Express and DynamoDB, but shows 2--3$\times$ higher write latencies.

\subsubsection{Choosing Storage for Data Processing}
Our results allow us to differentiate between the available serverless storage options. The option that provides the most economic scalable throughput is S3. Standard S3, however, offers the lowest out-of-the-box IOPS performance at the highest request latency. For these reasons, the recent S3 Express variant is an attractive alternative. S3 Express offers the highest IOPS throughput at consistent low latency, but at higher cost. DynamoDB provides the lowest latency, yet also the lowest throughput. Finally, EFS shows a balanced performance, but it is inferior to S3 Express in every evaluated dimension at a higher price point. We conclude that S3 is the most suited option for scalable data processing and focus the rest of our evaluation on methods based on object storage.

\subsection{Object Storage IOPS Scaling}
\label{section:storage_warming}

Object stores provide scalable throughput at request latencies that are acceptable for many analytical workloads \cite{Cai:2018, DBLP:journals/pvldb/TanGPYSDSAK19}. One major performance limitation of object stores is their low default IOPS, making them reject requests under spiking load. This is problematic for concurrent query workloads on the same datasets. Similarly, it is a challenge to serverless data analytics systems that shuffle intermediate data through object storage \cite{Pu:2019}. Every serverless worker node has to potentially read all relevant columns of all assigned partitions of all workers in the preceding stage. For terabyte-scale queries, even advanced shuffle strategies \cite{Mueller:2020, Perron:2020} require thousands of requests. In this experiment, we show how to scale up IOPS in S3 both reliably and efficiently. We also describe how S3 scales down partitions and IOPS when idle.

In the S3 object store, user data is horizontally partitioned on the object key namespace. Objects are organized in string key prefixes, which can span from an entire bucket to an individual object \cite{s3-prefixes}. The prefixes are backed by physical partitions on S3 storage nodes and serve 3.5K writes and 5.5K reads per second, respectively \cite{s3-practices}. To account for workload changes, prefixes are split and merged automatically and gradually over time (cf. \sectionref{section:serverless_storage}).

\subsubsection{IOPS Scaling}
\label{section:iops_scaling}
We examine the fraction of successful requests under carefully controlled increasing load to understand object storage IOPS scaling. This is necessary, because S3 throttles requests quickly when load spikes \cite{Pu:2019, s3-practices}.
For this experiment, we reuse the microbenchmark from the previous section with Lambda for compute and S3 for storage. The S3 client is configured with a request timeout of 200 ms for retries and exponential backoff \cite{Brooker:2019}. This results in an eager but not aggressive retry behavior. We start with 20 Lambda instances that each read one thousand 1 KB objects concurrently. An instance has four vCPUs and generates {\raise.17ex\hbox{$\scriptstyle\mathtt{\sim}$}}250-350 asynchronous requests per second, so the overall cluster saturates an S3 partition (with {\raise.17ex\hbox{$\scriptstyle\mathtt{\sim}$}}5--7K IOPS). We run 10 repetitions with this configuration. Then, we continue to run more configurations the same way, each incrementally adding two cluster instances (and {\raise.17ex\hbox{$\scriptstyle\mathtt{\sim}$}}600 IOPS load) up to a total of 100 instances and around 30K requests per second.

\begin{figure}[bt]
    \centering
    \includegraphics[width=1.0\linewidth]{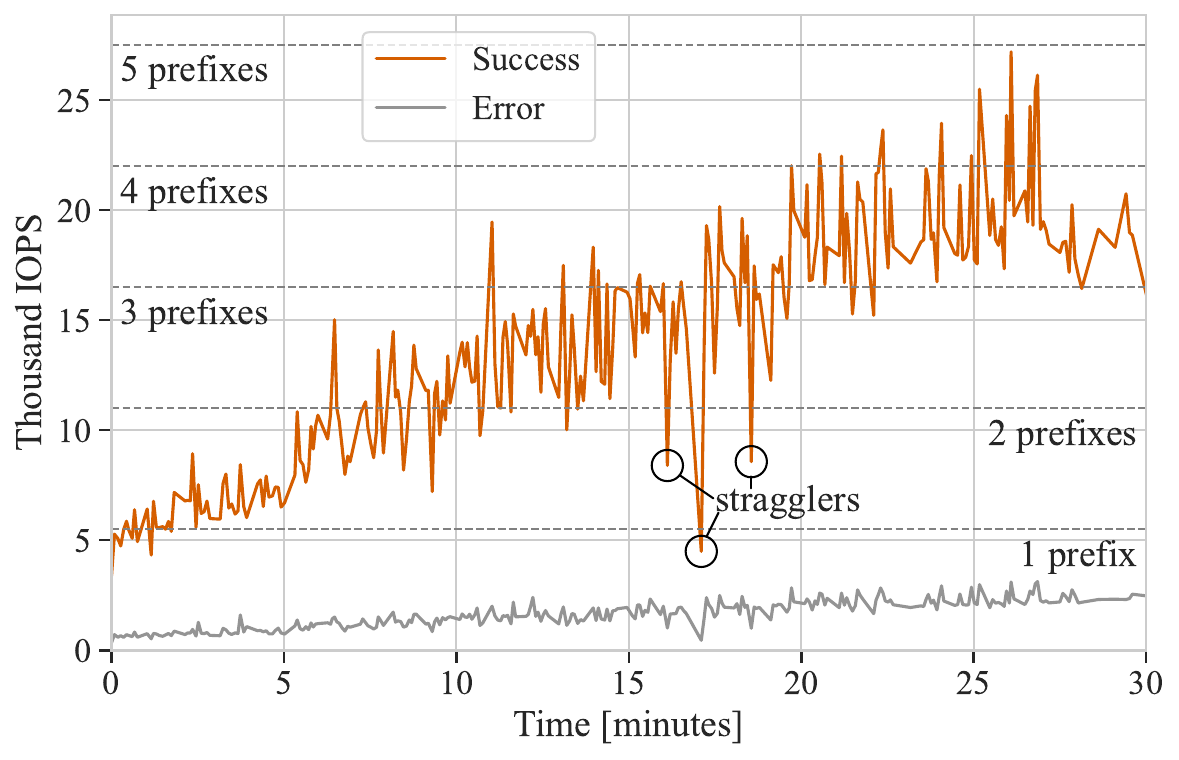}
    \vspace{-0.8cm}
    \caption{S3 IOPS scaling from one to five prefix partitions.}
    \vspace{-0.5cm}
    \label{figure:object_storage_warming}
\end{figure}

The results are depicted in \figureref{figure:object_storage_warming}. We plot the average successful and failed (throttled or timed out) read operations per second for each repetition over time. Our results show that S3 scales nearly linearly from {\raise.17ex\hbox{$\scriptstyle\mathtt{\sim}$}}5--27.5K IOPS
with this load pattern.
While scaling out, IOPS performance has a high variance with a relative standard deviation of up to 29\% for individual configurations. In addition, we observe three significant performance drops about 16--19 minutes into the experiment. Although the overall error rate is constant at just above 10\% throughout the experiment, few S3 clients see their requests repetitively being rejected. These clients then wait exponentially longer after every attempt and turn into stragglers in their respective repetition. Hence, the drops in IOPS are due to our client configuration and not S3's scaling behavior. We suspect that this is also the case for the previously mentioned study \cite{Pu:2019}.

\begin{figure}[b]
    \centering
    \includegraphics[width=1.0\linewidth]{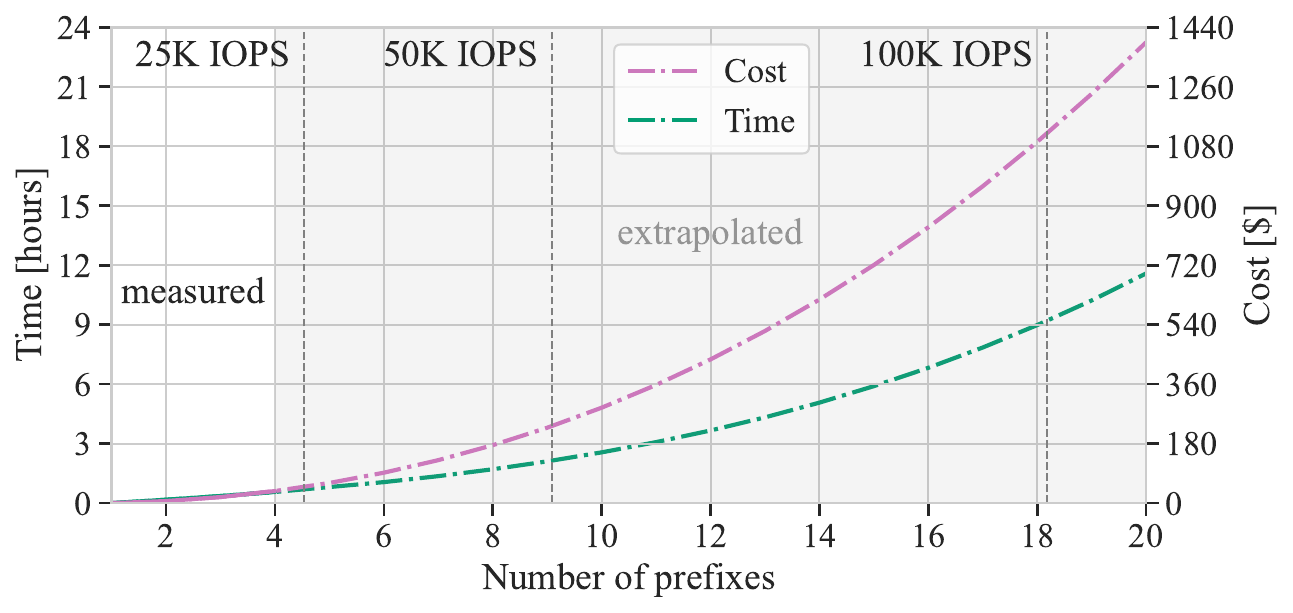}
    \vspace{-0.8cm}
    \caption{Required time and budget for S3 IOPS scaling.}
    \vspace{-0.5cm}
    \label{figure:warming_time_and_cost}
\end{figure}

In our experiment, we observe S3 scaling from a single partition serving 5.5K IOPS to five partitions providing 27.5K IOPS. This process takes about 26 minutes and 63 million requests costing \$25. To determine what it would take to make S3 partition our prefixes further for higher IOPS performance, we extrapolate time and cost based on our measurements. In \figureref{figure:warming_time_and_cost}, we show our measured and extrapolated data points for up to 20 prefix partitions totaling 110K IOPS. Given our load pattern and polynomial fitting method, we see that it would take about 2 hours and \$228 to reach 50K IOPS. Analogously, it would require 9 hours and \$1,094 to get 100K IOPS. This makes object storage IOPS scaling a quickly growing expense for users while S3 only allocates resources linearly and with delay as a form of admission control (cf. \sectionref{section:serverless_infrastructure}).

In experiments not shown for brevity, we further observe that prefix naming (e.g., prefixing object keys with hashes \cite{s3-practices, s3-stackoverflow}) does not impact IOPS scaling. Furthermore, sustained read load does not increase write IOPS performance. In fact, we are not able to scale write IOPS beyond a single partition's capacity (3.5K) with continuous write load (of up to 85 million requests over two hours).

\subsubsection{Downscaling Behavior}
\begin{figure}[bt]
    \centering
    \includegraphics[width=1.0\linewidth]{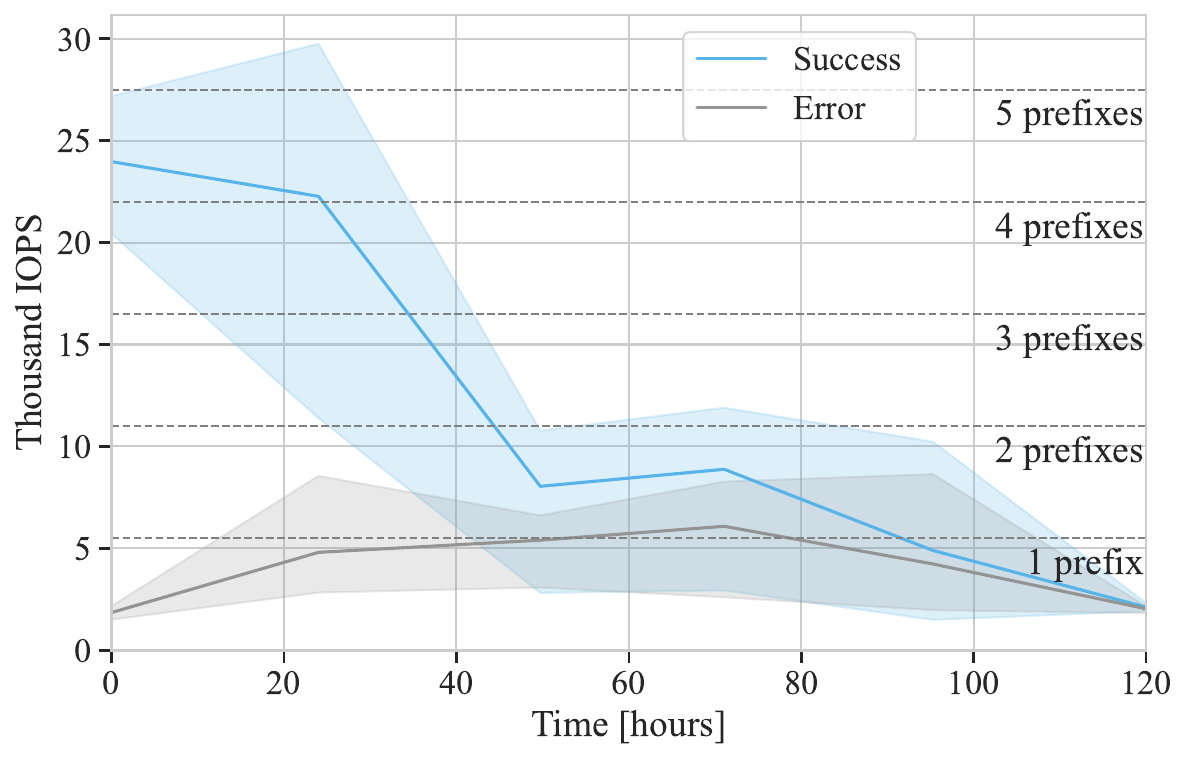}
    \vspace{-0.8cm}
    \caption{S3 scaling down from five to one prefix partitions under hourly and daily load patterns.}
    \vspace{-0.5cm}
    \label{figure:object_storage_cooling}
\end{figure}

After scaling up IOPS in S3, we study the process of scaling back down in periods of low load. This is to better understand when S3 begins to throttle requests and merge prefix partitions. We conduct this experiment in direct succession to the experiment in \sectionref{section:iops_scaling}. We first wait for an extended period of time before we run three repetitions of the last (and largest scale) employed configuration of the storage microbenchmark. We repeat this procedure until IOPS performance drops down to the level of a single partition and stays there. Since our experiment generates load against S3, it potentially influences its own outcome. There is a tradeoff between the frequency and accuracy of the measurements. For this reason, we run the experiment on two separately scaled buckets with hourly and daily measurement intervals, respectively. In \figureref{figure:object_storage_cooling}, we see the results of scaling down our S3 buckets from five partitions to one. We plot the band of IOPS across all three repetitions per interval for each series of measurements. We take the highest IOPS per interval as an indication for the number of the remaining partitions in the buckets. Our results indicate that the overall downscaling process takes between four and five days. After a full day of inactivity, all five partitions remain available to serve load. Two out of the five partitions continue to be available for an additional three days before IOPS performance returns to the level of a single partition. Since IOPS performance scales linearly and remains high over extended periods of low load, we conclude that IOPS scaling is a relevant optimization for analytical workloads, even if they are infrequent (with hourly or daily load patterns).

\begin{table}[b]
    \centering
    \caption{Datasets used in the experiments. Partition sizes are the mean Parquet file size with ZSTD compression.}
    \vspace{-0.3cm}
    \setlength\tabcolsep{5.5pt}
    \renewcommand{\arraystretch}{0.8}
    \begin{tabular}{lrrrr}
      \toprule
      \textbf{TPC Table} & \textbf{@ SF1000} & \multicolumn{2}{r}{\textbf{Partitions}} \\
                     & Size [GiB] & \# & Size [MiB]\\
      \midrule
        H-Lineitem & 177.4 & 996 & 182.4 \\
        H-Orders & 44.9 & 249 & 176.1 \\
      \midrule
        BB-Clickstreams & 94.9 & 1,000 & 92.7 \\
        BB-Item & 0.08 & 1 & 75.8 &  \\
      \bottomrule
    \end{tabular}
    \label{table:4_datasets}
    \vspace{-0.2cm}
\end{table}

\subsection{Exploiting Serverless Characteristics}
\label{section:exploiting_characteristics}
We now study how the described performance characteristics of serverless resources translate to data-intensive applications. Since the translation is complex, we first show the impact on data system components and then on full analytical queries. In the following experiments, we use queries and datasets from the TPC-H \cite{tpc-h} and TPCx-BB \cite{tpcx-bb} benchmarks. We run all queries on the tables of scale factor 1.000. The tables are partitioned into Parquet files and stored on S3. We employ the standard generators and do not  partition or sort on any specific keys. We provide details in \tableref{table:4_datasets}. We run the queries on the Skyrise query engine. Query workers have 4 vCPUs and 7.076 MiB RAM. Intermediates are stored in S3.

\subsubsection{Network Bursting for Scan-heavy Queries}
\label{section:component_benchmarks}

\begin{figure}[bt]
    \centering
    \includegraphics[width=1.0\linewidth]{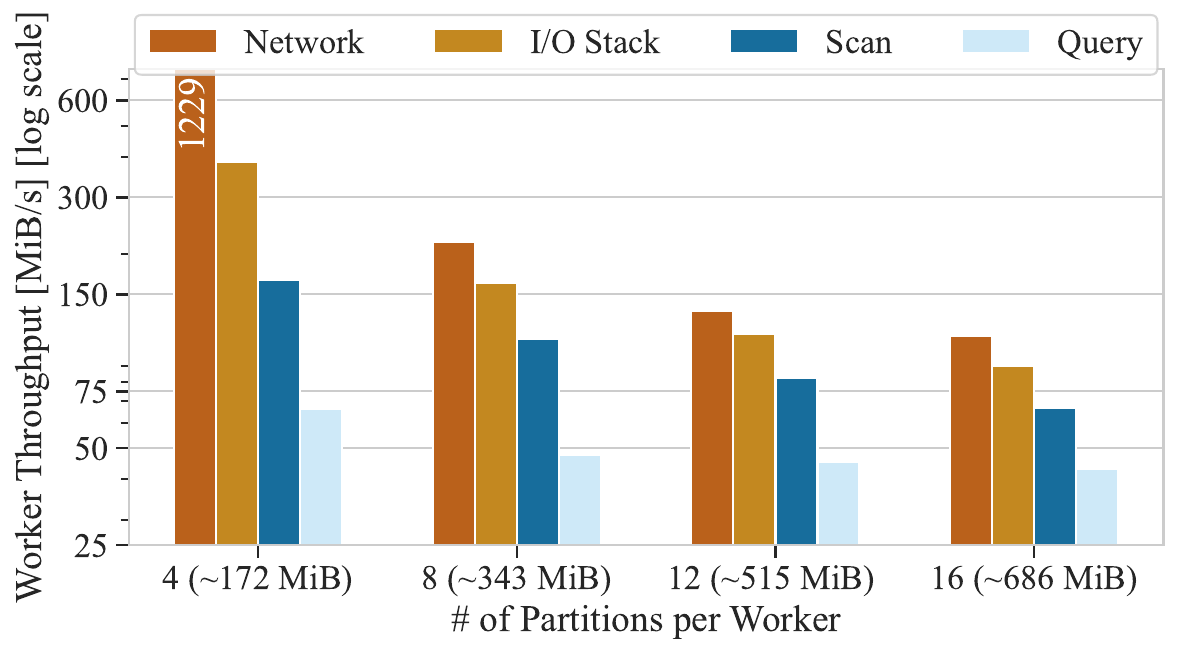}
    \vspace{-0.7cm}
    \caption{Query worker throughput for given input sizes within and beyond the network burst budget with TPC-H Q6.}
    \vspace{-0.3cm}
    \label{figure:scan_throughput}
\end{figure}

In our network analysis, we determine a budget of 300 MiB for unthrottled throughput, which benefits throughput-heavy operations like table scanning. We show the benefit of burst-awareness by running the scan-heavy TPC-H query Q6. We assign workers an increasing number of partitions, gradually exceeding their budgets. We present our results in \figureref{figure:scan_throughput}. We plot the expected throughput per worker according to our network model and the actual throughput of the Skyrise engine's I/O stack, the scan operator, and the complete query. We observe the performance impact of S3 request handling, decompression and deserialization, and the scan and full query logic. Across the partitioning settings, we see that queries fully exploiting the network burst are up to 53\% faster. We derive that serverless data systems benefit from calibrating and managing the network ingress/egress of their compute nodes.

\subsubsection{IOPS Scaling for Queries with Shuffles}
\label{section:application_benchmarks}

\begin{figure}[bt]
    \centering
    \includegraphics[width=1.0\linewidth]{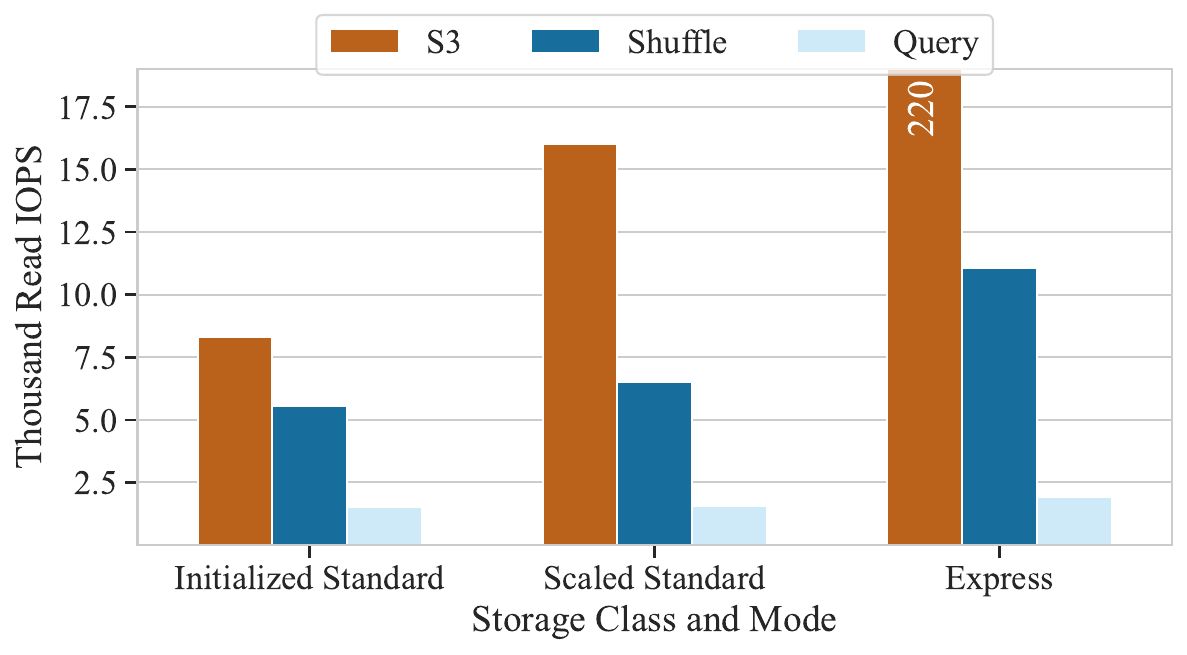}
    \vspace{-0.8cm}
    \caption{IOPS throughput of various S3 classes and modes and their performance impact on TPC-H Q12 and its shuffle.}
    \vspace{-0.6cm}
    \label{figure:shuffle_iops}
\end{figure}

Our analysis of object storage shows that IOPS performance scales under sustained load. We replicate and exploit this behavior at the query level. We run the I/O-heavy TPC-H Q12 join query with 320 workers. At this degree of parallelism, the shuffling for the join requires about 42.000 read operations and is constrained by default rate limiting. For shuffling, we employ three different storage setups. We use a new S3 Standard bucket, another bucket that has just been used for query execution for 15 minutes, and an S3 Express bucket. Our results are shown in \figureref{figure:shuffle_iops}. For reference, we plot the read IOPS throughput for all setups, as measured in \sectionsref{section:storage_throughput}{section:storage_iops}. We see that the runtime of the shuffle and the entire query are generally reduced by about 50\% and 20\%, respectively. While scaling IOPS of object storage takes too long to do as part of interactive queries, throughput should be considered when planning query parallelism.

\subsection{Query Performance Variability}
\label{section:performance_predictability}

Now that we better understand network bursting in serverless functions and object storage scaling, we quantify the impact of other local, geographical, and temporal aspects on the variance of overall query performance. For this set of experiments, we add two other queries that implement scan-heavy aggregation (TPC-H Q1) and I/O-bound MapReduce jobs (TPCx-BB Q3). We specifically employ these queries on synthetic data to avoid data and computational skew. We further ensure that the Skyrise query workers stay within their network burst budgets and consistently use either cold or warm function instances and storage buckets. We deploy and run our query suite in the AWS regions us-east-1, eu-west-1, and ap-northeast-1. We conduct one experiment by repetitively running our query suite
over a workday with 15 minute intervals between runs, dubbed cold. We then carry out a second experiment with the queries run back-to-back with no wait, i.e. warm, over three hours.

We present our results in \tableref{table:4_performance_variability}. We report two metrics, namely median to base median ratio (MR) and coefficient of variation (CoV). MR normalizes the query suite runtime with the median of the us-east-1 (US) region. We use CoV \cite{Schad:2010} as a measure of variation within a region. We see a mixed picture. The variance is low across the US and AP regions, but significant compared to the EU ({\raise.17ex\hbox{$\scriptstyle\mathtt{\sim}$}}50\%). In the EU, the startup of large function clusters takes significantly longer, likely due to contention within the region. For the US and AP regions, the local variability is higher, with the cold experiment showing yet higher variance than the warm one. We deduce that more frequent usage leads to pre-provisioning of resources and more robustness. Localized factors continue to considerably impact variability in performance, necessitating further examination.

\begin{table}[bt]
    \centering
    \caption{Performance variability between and within regions in short experiments and over a weekday.}
    \vspace{-0.2cm}
    \setlength\tabcolsep{5.5pt}
    \renewcommand{\arraystretch}{0.8}
    \begin{tabular}{lrrrr}
      \toprule
      \textbf{Measure} & \textbf{US} & \textbf{EU} & \textbf{AP} \\
      \midrule
        Cold MR (US) & 1 & 1.48 & 0.95 \\
        Cold CoV (24h) & 22.65 & 4.76 & 7.65 \\
      \midrule
        Warm MR (US) & 1 & 1.52 & 0.96 \\
        Warm CoV (3h) & 5.23 & 8.96 & 6.44 \\
      \bottomrule
    \end{tabular}
    \label{table:4_performance_variability}
    \vspace{-0.5cm}
\end{table}

\section{Economic Viability}
\label{section:economic_viability}

In this section, we discuss the economic implications of building data processing systems on serverless infrastructure. We explain the assumptions for our discussion in \sectionref{section:assumptions}. We then examine the cost-saving potential of FaaS-based query execution in \sectionref{section:compute_tradeoff}. In \sectionref{section:storage_tradeoff}, we determine the break-even points for scanning and shuffling data on serverless storage.

\subsection{System Architecture and Cloud Pricing}
\label{section:assumptions}

In our examination, we assume a distributed system architecture that disaggregates both persistent and ephemeral storage and uses small and stateless compute nodes. This is the inherent architecture of serverless data processors \cite{Mueller:2020, Perron:2020} and also the target design for some commercial cloud systems \cite{Melnik:2020, Vuppalapati:2020}. While serverless systems use this architecture due to the restrictions discussed in \sectionref{section:serverless_infrastructure}, industrial systems adopt it for elasticity. Making this assumption, we factor out the inefficiencies of disaggregation and distribution \cite{Kanev:2015, Seemakhupt:2023} compared to monolithic and single-node systems, which can potentially cache and process data entirely in memory.

Beyond this, we assume the current service pricing models of AWS (cf. \tablesref{table:2_compute_services}{table:2_storage_services}, \cite{ebs-pricing}). They are comparable to the ones from Azure \cite{azure-pricing} and GCP \cite{gcp-pricing} and are reasonably stable over time \cite{aws-pricing-stability}.

\subsection{Breaking Even with Serverless Compute}
\label{section:compute_tradeoff}

FaaS platforms have higher compute unit prices and performance overheads \cite{Scheuner:2020} than conventional IaaS platforms. In return, they offer automatic, elastic, and fine-grained scalability. In this section, we study the economic tradeoff of IaaS and FaaS deployment of data processing systems. We determine the performance overhead and cost of FaaS-based execution for selected queries. In addition, we identify the cost-savings potential enabled by elasticity.

\begin{table}[bt]
    \centering
    \caption{Execution statistics and derived economic metrics: Break-even FaaS query throughput for peak-provisioned IaaS and intra-query peak-to-average node ratio.}
    \vspace{-0.3cm}
    \setlength\tabcolsep{8.0pt}
    \renewcommand{\arraystretch}{0.8}
    \begin{tabular}{lrrrr}
      \toprule
      \textbf{Query} & H-Q6 & H-Q12 \\
      \midrule
        IaaS Runtime [s] & 5.2 & 18.1 \\
      \midrule
        FaaS Runtime [s] &  5.7 & 19.2 \\
        Cumulated Time [s] & 515.9 & 2,227.3 \\
        FaaS Cost [\textcent] & 4.87 & 21.19 \\
        \textit{Break-Even [Q/h]} & \textit{558} & \textit{128} \\
        \textit{Peak-to-Average-Nodes} & \textit{2.21}$\times$ & \textit{2.43}$\times$ \\
        \midrule
        Storage Requests & 1,401 & 30,033 \\
        Shuffle I/O Size [KiB] & 0.4 & 1.1--2,078 \\
        Storage Cost [\textcent] & 0.16 & 1.39 \\
      \bottomrule
    \end{tabular}
    \label{table:5_compute_tradeoff}
    \vspace{-0.5cm}
\end{table}

For this experiment, we rerun our query suite from \sectionref{section:performance_predictability} in two configurations. We first run the queries on Skyrise in Lambda as before with each function having 4 vCPUs and 7.076 MiB RAM. Then, we deploy Skyrise on a cluster of 284 EC2 C6g.xlarge VMs with 4 vCPUs and 8 GiB RAM. The functions are warmed up and the VMs are started before the experiment begins. For both configurations, the query plans and physical resources are the same. The Skyrise workers employ S3 to read the base tables and shuffle intermediate results. We run the query suite ten times each and collect statistics from the run with the median runtime. The statistics include the runtime, the accumulated function lifetime, and the number and size of the storage requests per query. We present our results for the queries TPC-H Q6 and Q12 in \tableref{table:5_compute_tradeoff}.

\textbf{Query Runtime Slowdown.} In the FaaS deployment of Skyrise, the end-to-end latencies for Q6 and Q12 are 10\% and 6\% higher. The primary reason is the startup time of the functions for every query stage compared to no startup overhead in the IaaS deployment with pre-provisioned VMs. In addition, there are occasional cold start stragglers, in particular for the coordinator. These stragglers do not impact cost, because other functions do not idle waiting for them. 

\textbf{Query Cost and Break-Even Throughput.} We calculate the FaaS cost of a query based on the aggregated lifetimes of both the coordinator and worker functions in all stages. We relate the query cost to the cost of a peak-provisioned VM cluster to determine the break-even throughput. A C6g.xlarge instance costs 0.136 \$/h. The peak number of instances used for Q6 is 201 and for Q12 is 284. Thus, FaaS deployment is economical for up to 558 runs per hour of Q6 or 128 runs of Q12. For adaptively provisioned clusters with higher utilization, the break-even throughput decreases proportionally.

\textbf{Intra-Query Elasticity.} Analytical queries  consist of multiple stages that may have very different input sizes and computational requirements. Skyrise schedules 284 nodes in the first stage of Q12 to scan and filter 222.3 GiB and a single node to aggregate 105.5 KiB (six orders of magnitude smaller) into the final result in the last stage. We calculate the peak-to-average node ratio across stages as potential cost-savings factor compared to static peak node provisioning for queries. For Q12, this ratio is 2.43$\times$.

\subsection{Breaking Even with Serverless Storage}
\label{section:storage_tradeoff}

Cloud functions can neither cache data beyond their short lifetimes nor can they communicate directly. Thus, FaaS-based query workers need to access remote cloud storage to scan and exchange data. This section investigates the cost implications of these limitations. In our discussion, we exclude performance concerns and assume scans overlap with computation and shuffles are throughput-bound.

\subsubsection{Reads in the Cloud Storage Hierarchy} We base our argument on caching on Gray’s regularly revisited five-minute rule \cite{Gray:1987, Appuswamy:2017} for trading off memory and disk accesses. We introduce two variations of the rule to account for different cloud storage pricing models. We use the first variant for cloud storage that is priced by capacity only, such as VM-based RAM and SSDs, as well as network drives. We calculate the break-even interval (BEI) in seconds as follows.
\begin{align*}
     BEI = \frac{PagesPerMB}{AccessesPerSecondPerDisk} \times \frac{RentPerHourPerDisk}{RentPerHourPerMBofRAM} \enspace
\end{align*}

The second variant reflects pricing by the number of requests, as in serverless object storage and key-value stores.

\vspace{-3mm}

\begin{align*}
     BEI &= PagesPerMB \times \frac{PricePerAccessToTier2}{RentPerSecondPerMBofTier1} \enspace
\end{align*}

\begin{table}[bt]
    \centering
    \caption{Break-even intervals for different data access sizes and storage combinations in AWS (us-east-1) in July 2024.}
    \vspace{-0.3cm}
    \setlength\tabcolsep{8.0pt}
    \renewcommand{\arraystretch}{0.8}
    \begin{tabular}{lrrrr}
      \toprule
      \textbf{Access Size} & 4 KiB & 16 KiB & 4 MiB & 16 MiB \\
      \midrule
        RAM/SSD & 38s & 31s & 31s & 31s \\
        RAM/EBS & 27min & 7min & 3min & 3min \\
        RAM/S3 Standard & 2d & 12h & 3min & 41s \\
        RAM/S3 Express & 23h & 6h & 36min & 39min \\
      \midrule
        SSD/S3 Standard & 59d & 15d & 1h & 21min \\
        SSD/S3 Express & 29d & 7d & 18h & 20h \\
        SSD/S3 X-Region & 70d & 26d & 11d & 11d  \\
      \bottomrule
    \end{tabular}
    \label{table:5_scan_tradeoff}
    \vspace{-0.3cm}
\end{table}

For our comparison along the hierarchy of cloud storage options for query workers, we include RAM, SSDs, EBS network drives \cite{ebs-specification}, and S3 object storage. We assume workers to run on  EC2 C6gd VMs with NVMe SSDs \cite{ec2-specification}. We further assume on-demand prices. Lower reserved prices increase the break-even intervals proportionally, whereas higher Lambda prices decrease them. \tableref{table:5_scan_tradeoff} shows the results of our calculations. We draw the following conclusions.

\textbf{Relevance of SSDs.} The break-even for RAM and SSD using 4 KiB accesses is 38s, one order of magnitude less than a decade ago \cite{Appuswamy:2017}. This is due to increased IOPS performance and decreased prices. The interval for larger accesses does not get much shorter because the maximum SSD bandwidth in EC2 of 2 GiB/s \cite{ec2-specification, Leis:2024} becomes the bottleneck, limiting the SSD IOPS in above formula. Conversely, the break-even for SSDs and object storage is hours to days for all but very large ($\ge$16 MiB) accesses. As a result, caching data on SSDs is economical for a wide range of access sizes and frequencies. While cloud functions support SSDs, their cache lifetime is limited to their own. They do not support network drives.

\textbf{Analytics on Cold Data.} Query workers do not benefit from SSD caches when accesses are in the megabyte-scale and they occur at most on an hourly basis, e.g., one 4 MiB access per hour. This is the definition of cold data that serverless systems target \cite{Mueller:2020}.

\textbf{Pricing Model.} The RAM/SSD break-even intervals are constant within an EC2 instance family (e.g., C6g). This is due to SSD IOPS performance growing with its instance and price \cite{ec2-specification}, and both parameters being on opposite sides of above equation. Data transfer fees, as for S3 Express and cross-region access, invalidate the initial rule that the break-even is inversely proportional to the access size.

\begin{table}[bt]
    \centering
    \caption{Break-even data access sizes for different instance types and storage systems in AWS (us-east-1) in July 2024.}
    \vspace{-0.3cm}
    \setlength\tabcolsep{3.0pt}
    \renewcommand{\arraystretch}{0.8}
    \begin{tabular}{lcccc}
      \toprule
      \textbf{Instance} & C6g.xlarge & C6g.8xlarge & C6gn.xlarge & C6gn.xlarge \\
      \textbf{Type} & on-demand & on-demand & on-demand & reserved \\
      \midrule
        S3 Standard & 2 MiB & 2 MiB & 7 MiB & 16 MiB \\
        S3 Express & -- & -- & -- & -- \\
      \bottomrule
    \end{tabular}
    \label{table:5_shuffle_tradeoff}
    \vspace{-0.5cm}
\end{table}

\subsubsection{Impact of Shuffle I/O Size} To shuffle intermediate results in a serverless system, every worker reads its respective partition(s) from every object of the preceding stage(s) from object storage. For large queries, the cost of the resulting read requests dominate the overall query cost. For this reason, many systems employ key-value stores on provisioned VM clusters \cite{Klimovic:2018, Pu:2019, Perron:2023} to shuffle intermediates. The shuffle capacity of a cluster is the aggregated network throughput of the VMs and its cost is the combined cost of the VMs. Since object storage requests are priced independently of their size, there is a break-even access size at which object storage becomes more economical for shuffling. We calculate this access size (BEAS) in MB as follows.
\begin{align*}
     BEAS &= PricePerAccess \times \frac{MBPerHourPerServer}{RentPerHourPerServer} \enspace
\end{align*}

\tableref{table:5_shuffle_tradeoff} shows our results for different cluster VM types and the S3 Standard and Express storage classes. The VM types are from the C6g instance family, including the network-optimized C6gn variant with four times the network throughput at both on-premise and reserved pricing \cite{ec2-pricing}. We derive the following insights.

\textbf{Case for Large Accesses.} Object storage is the cheaper shuffle medium when average access sizes are larger than 2 -- 16 MiB, depending on the VM type and pricing model. In distributed query execution, intermediates are highly partitioned and individual I/Os tend to be small. In our query experiments, they are {\raise.17ex\hbox{$\scriptstyle\mathtt{\sim}$}}1 KiB -- 2 MiB (cf. \tableref{table:5_compute_tradeoff}). There, however, is a range of techniques to increase I/O sizes, including write combining and staged shuffling \cite{Mueller:2020, Perron:2020}.

\textbf{Pricing Model.} The break-even access sizes are constant within EC2 VM families, since network throughput grows proportionally with VM size and price. The S3 Express storage class never breaks even with VM clusters due to its data transfer cost component.

\section{Discussion}
\label{section:discussion}

In this section, we present key takeaways from our evaluation of the performance and cost efficiency of serverless cloud infrastructure.

\textbf{Serverless Performance.}
Our results show that the key techniques of serverless systems (e.g., tenant isolation and placement, as well as rate limiting) have a large impact on performance. We identify four aspects that have not yet been studied in detail.
\begin{enumerate}
    \item \textbf{Network rate limiting:} We show that serverless functions and VMs are subject to network bandwidth limiting once they consumed a deterministic burst capacity. Thus, their ingress/egress should be aligned to maximize bandwidth. The accelerated bandwidth can benefit scan-heavy queries.
    \item \textbf{Storage IOPS scaling:} Object storage is the most suited option for serverless data analysis.
    We demonstrate that the strict request rates of object storage deterministically increase under sustained load and decrease in extended idle periods. This process can be used to optimize shuffle queries.
    \item \textbf{Variability factors:} Regional variance can be substantial, but generally temporal variance is higher. More frequent usage leads to pre-provisioning of resources and more robustness. Considerable sources of local variability remain for further investigation.
    \item \textbf{Security conflicts:} Virtual network partitions (\eg VPCs used in industry) currently hinder the elasticity of serverless networks.
\end{enumerate}

\textbf{Serverless Economics.}
We observe that unit prices are higher in serverless systems, which provision resources on behalf of the users to provide elasticity. We see three ways to optimize cost.
\begin{enumerate}
    \item \textbf{Infrequent and peak usage:} Users only pay for consumed resources and benefit when workloads are infrequent or peak unpredictably. So the serverless pricing model should be combined with the cost models for provisioned resources.
    \item \textbf{Intra-job elasticity:} Analytical queries and machine learning pipelines have varying resource demands and can benefit from intra-job elasticity enabled by serverless compute.
    \item \textbf{Economic data tiering:} Our examination suggests that cold (hourly accessed) data should be kept in object storage and accessed in MiB sizes. Warmer data stays in VM-based SSDs.
\end{enumerate}

\textbf{Transaction Processing.} The request latencies and prices of current serverless storage services are generally inadequate for fine-grained operational workloads.

\textbf{Generality of Results.} We believe that our results have a wide relevance, since the studied performance effects are present in VMs and object storage, which are the major building blocks for modern commercial data analysis systems. These systems further start to adopt serverless compute resources for functionality, such as UDFs and ETL. We acknowledge that concrete numbers may change between major cloud providers, over time, and between different geographies. We offer our open-source tooling and methods to validate and revalidate our results.

\section{Related Work}
\label{section:related_work}

This section summarizes related work on serverless infrastructure and data processing systems.

\textbf{Analysis of Serverless Infrastructure.}
Prior work includes benchmark frameworks and studies for serverless systems from all major cloud providers \cite{Wang:2018, Maissen:2020, Pons:2021, Scheuner:2020}. On the application level, they focus on web, IoT, and media applications that require little coordination and state \cite{Kim:2019, Yu:2020, Copik:2021, Grambow:2021}. On the resource level, they cover aspects including the performance and isolation of cloud function CPUs, memory, and disk storage. They provide insights into FaaS platform overheads and scalability. There has been evaluation of the network characteristics of VMs \cite{Schad:2010, Uta:2020} and cloud functions \cite{Mueller:2020, Wawrzoniak:2021}. Serverless storage has been studied in \cite{Klimovic:2018, Pu:2019, Roy:2021, Palepu:2022, Durner:2023}. Another body of work focuses on performance variability in IaaS \cite{Schad:2010, Uta:2020} and FaaS platforms \cite{Perron:2020, Ustiugov:2021}.

We instead evaluate the performance of serverless resources for large-scale and stateful applications. We perform a detailed analysis in AWS running millions of cloud functions and billions of storage requests moving hundreds of terabytes of data. We characterize the burstable network performance of serverless functions and the scalability of various serverless storage systems, including the recent S3 Express. We demonstrate that these properties translate to application performance and quantify the remaining sources of performance variability. We present cost break-even points for serverless compute and storage in the data processing context.

\textbf{Serverless Data Systems and Applications.} Recently, there have been several system prototypes to explore the viability of serverless infrastructure for data processing. Some support general-purpose, MapReduce-style processing
\cite{corral, Jonas:2017, Kim:2018, Pu:2019, Jarachanthan:2022, Mahling:2023}, and some are SQL query execution engines \cite{Mueller:2020, Perron:2020, Bian:2023, Perron:2023}. Other works evaluate serverless resources for different I/O-intensive workloads, such as machine learning training \cite{Carreira:2019, Jiang:2024}. Most systems are closed-source, and none allow for the analysis of the impact of resource-level properties on system components and full queries.

In contrast, Skyrise is open-source and allows to explore serverless infrastructure properties across the stack with a suite of micro-benchmarks and an integrated serverless query engine.

\section{Conclusion}
\label{section:conclusion}

We perform an in-depth analysis of serverless network, storage, and compute behavior for data processing in an extensive series of large-scale cloud experiments. Our results provide a detailed understanding of network performance bursting and I/O warming and their influence on query processing. Using our analysis framework, Skyrise, we execute full queries and compare serverless and VM-based execution. We derive several cost break-even points to determine when serverless query processing is economical. 

\begin{acks}
We thank the Master's students that worked on the Skyrise project and implemented components of the evaluation framework: Lars Jonas Bollmeier, Fabian Engel, Tobias Maltenberger, and many more.
This work was partially funded by SAP, the AWS Cloud Credit for Research Program, the German Research Foundation (414984028), and the European Union’s Horizon 2020 research and innovation programme (957407).
\end{acks}

\balance
\bibliographystyle{ACM-Reference-Format}
\bibliography{bibliography}

\end{document}